\documentclass[journal]{IEEEtran} 

\usepackage{cite}

\ifCLASSINFOpdf
  \usepackage[pdftex]{graphicx}

  \DeclareGraphicsExtensions{.pdf,.jpeg,.png}
\else
   \usepackage[dvips]{graphicx}

   \DeclareGraphicsExtensions{.eps}
\fi

\usepackage[cmex10]{amsmath}

\usepackage{array}
\usepackage{arydshln}

\usepackage{eqparbox}

\usepackage[caption=false,font=footnotesize]{subfig}

\usepackage{amsfonts}
\usepackage{amssymb}
\usepackage{balance}
\usepackage{enumerate}
\usepackage{latexsym}
\pdfoutput=1
\usepackage{hyperref}
\hypersetup{
    colorlinks=true,
    linkcolor=black,
    citecolor=black,
    filecolor=black,
    urlcolor=black,
}

\usepackage{floatrow}

\usepackage[cmex10]{amsmath}
\usepackage{mathrsfs}
\usepackage{multirow}

\usepackage{leftidx}
\usepackage{stmaryrd}
\usepackage{dsfont}

\usepackage{algorithm}
\usepackage{algorithmic}

\usepackage{hhline}
\newcommand{\tens}[1]{\boldsymbol{\mathcal{#1}}}
\newcommand{\tenselem}[1]{\mathcal{#1}}

\newcommand{\matr}[1]{\boldsymbol{#1}}

\newcommand{\vect}[1]{\boldsymbol{#1}}

\newcommand{\khatri}{\odot}     
\newcommand{\hadam}{\boxdot}    
\newcommand{\kron}{\otimes}     
\newcommand{\T}{{\sf T}}        
\newcommand{\leftinv}[1]{\leftidx{^\dagger}\!{\matr{#1}}}
\makeatletter
\newcommand{\diag}[1]{\mathop{\operator@font diag}\{#1\}}
\renewcommand{\vec}{\mathop{\operator@font vec}}   %
\newcommand{\vecT}{\mathop{\operator@font vecT}}   %
\newcommand{\Unvec}{\mathop{\operator@font Unvec}}   %
\newcommand{\Span}{\mathop{\operator@font span}}
\makeatother

\makeatletter
\renewcommand{\maketag@@@}[1]{\hbox{\m@th\normalsize\normalfont#1}}%
\makeatother

\begin{document}
%
\title{Exploring multimodal data fusion through joint decompositions with flexible couplings}

\author{Rodrigo~Cabral Farias*,%
        ~Jeremy~Emile~Cohen~%
        and~Pierre~Comon,~\IEEEmembership{Fellow,~IEEE}
\thanks{The authors are with CNRS, Gipsa-Lab, University of Grenoble Alpes, F-38000 Grenoble, France (e-mail: firstname.lastname@gipsa-lab.fr). This research was supported by the ERC Grant AdG-2013-320594 ``DECODA'', a part of which  was submitted to LVA/ICA 2015 \cite{Farias2015}. }

}


\maketitle

\begin{abstract}
A Bayesian framework is proposed to define flexible coupling models for joint tensor decompositions of multiple data sets. Under this framework, a natural formulation of the data fusion problem is to cast it in terms of a joint maximum \textit{a posteriori} (MAP) estimator. Data driven scenarios of joint posterior distributions are provided, including general Gaussian priors and non Gaussian coupling priors. We present and discuss implementation issues of algorithms used to obtain the joint MAP estimator. We also show how this framework can be adapted to tackle the problem of joint decompositions of large datasets. In the case of a conditional Gaussian coupling with a linear transformation, we give theoretical bounds on the data fusion performance using the Bayesian Cram\'er-Rao bound. Simulations are reported for hybrid coupling models ranging from simple additive Gaussian models, to Gamma-type models with positive variables and to the coupling of data sets which are inherently of different size due to different resolution of the measurement devices.

\end{abstract}


%
\IEEEpeerreviewmaketitle

\section{Introduction}\label{intro-sec}

In some domains such as neurosciences, metabolomics and data sciences, various data gathering devices are used to collect information on some underlying phenomena. Since it may occur that no data set contains a complete view of the phenomena, data fusion may be used to blend the different views provided by the complementary data sets, thus allowing a more comprehensive view. It is not then surprising that multimodal data fusion, \textit{i.e.} fusion of heterogeneous data, has become an important topic of research in these domains \cite{Sui2012,Acar2013a,Ermics2012}.

One way of defining a framework for multimodal data fusion is to cast it as a problem of latent variable analysis. Variations of the hidden variables are assumed to explain most of the variations in the measured data sets. Because the data sets are considered to be different views of the same phenomena, we suppose that the latent models are coupled through subsets of their variables. Here, coupling means statistical dependence between these subsets of variables. By exploiting this coupling in the joint estimation of the latent models, we expect that the information from one data set will improve the estimation accuracy and interpretability of the latent variables related to the other data set.

An early example of the framework explained above dates back to Hotelling's canonical correlation analysis (CCA) \cite{Hotelling1936}. In CCA a pair of subspaces is obtained such that the projections of the data vectors on their corresponding subspaces are maximally correlated. It has been shown \cite{Bach2005} that such an approach is equivalent to estimating two linear transformations of a shared latent white Gaussian vector from noisy measurements. While the linear transformations link the information in the data vectors within each data set, the underlying latent Gaussian vector allows the coupling between data sets.

In a deterministic setting, coupled (or linked) factorization models for multiple multiway ($ N $-dimensional) datasets were first described in \cite{Harshman1984} and they were repopularized recently in data sciences in \cite{Singh2008}. The problem of joint nonnegative matrix factorization was considered in \cite{Singh2008} under the constraint that one of the factors is shared by all matrices. In both cases, the coupling occurs through equality constraints on latent factors. Following these works, the framework of coupled tensor factorization was revisited in \cite{Banerjee2007,Lin2009}. Variations on this framework, such as tensor-matrix factorizations \cite{Acar2013b,Acar2011} and more general latent models \cite{Sorber2015} have also been considered. Uniqueness properties and the development of algorithms for the exact coupled tensor decomposition problem are proposed in \cite{Sorensen2015a} and \cite{Sorensen2015b}, while algorithms for the coupled tensor approximation problem under general cost functions are developed in \cite{Yilmaz2011,Ermics2012,Csimcsekli2015}. More flexible models for the couplings have been proposed in \cite{Acar2013a}. Instead of considering equality constraints for the entire factors of a tensor model, only a few components or elements are constrained. In \cite{Seichepine2014} the problem of coupled nonnegative matrix factorization is considered and a flexible coupling is proposed by assuming that the shared components are similar in $\ell_{1}$ or $\ell_{2}$ sense. Recently, factors coupled through their numerical derivatives have been considered \cite{Rivet2015}, which is a generalization of \cite{Acar2013b}, but still a particular case of this work.

In this article, we propose a generalization of the above flexible models for joint decompositions using a Bayesian estimation approach. This allows modeling approximate couplings and the derivation of performance bounds.

\section*{Outline and contributions}
\begin{itemize}
\item After introducing our Bayesian framework, we present two general examples of coupling priors such as joint Gaussian priors and non Gaussian conditional distributions. We describe  two algorithms for joint tensor approximate decomposition, a modified least squares algorithm for the joint Gaussian model and a multiplicative update for a positive coupling model.

In contrast to \cite{Csimcsekli2015}, we focus on stochastic models for the coupling between the factors in the decompositions and not on separate stochastic models for each factor. As the model considered in \cite{Seichepine2014}, our model is more flexible than the coupled model proposed in \cite{Acar2013a}, since the coupled variables do not need to be equal. Moreover, the joint Gaussian model and hybrid model that we propose are more general than similarity between factors, since our model allows different linear transformations on the coupled factors. This is useful  if we want to fuse data sets which are measured in different domains, for instance one in time and the other in frequency, or which are sampled differently. 

\item We discuss implementation issues related to this flexible coupling approach. In the case of a simple Gaussian coupling model, we propose a joint compression approach to deal with large data sets. To our knowledge, our work seems to be the first to discuss and propose a solution for the approximation of large coupled decompositions.

\item Since we define the coupled approximation problem with a Bayesian framework, we are able to obtain bounds on approximation variance in terms of mean squared error (MSE) using Cram\'er-Rao bounds (CRB). In the case where all parameters are random and both measurements and factors are Gaussian, we show how to obtain the Bayesian Cram\'er-Rao bound (BCRB) for a coupled tensor model, while in the case where only the coupling is random, we show how to obtain the Hybrid Cram\'er-Rao bound (HCRB).

Although CRB for tensor models have been developed in \cite{Liu2001} and \cite{TichPK13:tsp}, this article is the first to present bounds on the coupled tensor model.

\item We simulate the proposed algorithms to retrieve the factors of two tensor models coupled through one of the factors. The simulations are carried out for several types of coupling, including Gamma-type coupling of positive variables, and coupling of data sets which are inherently of different size due to different time resolution of the measurement devices.

Gamma-type measurements can be dealt with the framework presented in \cite{Csimcsekli2015}, but not Gamma couplings as presented here. Similarly, the coupling of data sets with different sizes cannot be handled with the flexible models presented in \cite{Acar2013a} or \cite{Seichepine2014}. Note also that this type of model appears naturally in multimodal data fusion since nothing guarantees that measurement devices of different nature will generate data sets with the same resolution. Our approach in this case can be seen as a generalization of the pan-sharpening approaches considered in remote sensing \cite{Yokoya2012}. 

\end{itemize}

\section*{Notation}
To simplify the presentation, all quantities will be assumed to belong to the real field. 
Scalars (resp. vectors) are denoted by lower case, \textit{e.g.} $x$, (resp. bold lower case \textit{e.g.} $\vect{x}$) letters. Matrices are denoted by upper case bold letters, \textit{e.g.} $\matr{X}$, while tensors by calligraphic letters, \textit{e.g.} $\tens{X}$. Elements of a given array are indicated by subscripts and are non-bold, \textit{e.g.} $\tenselem{X}_{ijk}$. 
Vectorization of matrices is indicated by $ \vec(\cdot) $, and stores bijectively all entries of an array into a column vector column-wise \cite{Brew78:cs}. The Kronecker product of two matrices $ \matr{X} $ and $ \matr{Y} $ is denoted by $ \matr{X}\kron \matr{Y} $, while the Khatri-Rao product (column-wise Kronecker product) by $ \matr{X}\khatri \matr{Y} $ \cite{Brew78:cs}. The Hadamard (element-wise) product, division and ($x$-th) power of matrices are denoted by $ \hadam $, $ \boxslash $ and $ \left[\cdot\right]^{x}$. 
 The left inverse of a full column-rank matrix $\matr{M}$ is denoted by $\leftinv{M}$, and defined as $\leftinv{M}=(\matr{M}^\T\matr{M})^{-1}\matr{M}^\T$. The right inverse of a full row-rank matrix is $\matr{M}^{\dagger}=\matr{M}^\T(\matr{M}\matr{M}^{\T})^{-1}$. Both are actually computed via the \emph{economy size} singular value decomposition (SVD) in practice.
We denote the trace of a matrix by $ \text{tr}(\cdot) $. 
A vertical stack of matrices or vectors will be denoted with semi columns as $[\vect{x} ; \vect{y}]$. A multilinear product of matrices $\matr{A}$, $\matr{B}$ and $\matr{C}$ with same number of columns is denoted by $(\matr{A} , \matr{B}, \matr{C})$; in other words,  it is the 3-way array defined by $(A, B, C)_{ijk} = \sum_r A_{ir} B_{jr} C_{kr}$. 
$\matr{Y}_{(i)}$ denotes the $i$-th matrix unfolding\footnote{Also referred to as ``mode-$i$ matricization.''} of tensor $\tens{Y}$, which is obtained by stacking all matrix slices along mode $i$; see a detailed definition in \textit{e.g.} \cite{Kolda2009} or \cite{Comon2014}. In particular, if $\tens{Y} =(\matr{A}, \matr{B}, \matr{C})$, then $\matr{Y}_{(3)}=\matr{C}(\matr{B}\khatri\matr{A})^\T$, and if  $\tens{Y}$ is of size $I\times J\times K$, then $\matr{Y}_{(3)}$ is $K\times IJ$. 


\section{Coupled decompositions: a Bayesian approach}
Consider two arrays of noisy measurements, $ \tens{Y} $ and $ \tens{Y}' $, which can be tensors of possibly different orders and dimensions. Arrays $ \tens{Y} $ and $ \tens{Y}' $ are related to two parametric models characterized by parameter vectors $ \vect{\theta} $ and $ \vect{\theta}' $, respectively.

For instance, if $ \tens{Y} $ is a matrix (a second order tensor) to be diagonalized, the model can be the SVD $ \tens{Y}=\matr{U} \matr{\Sigma} \matr{V}^{\T}+\matr{E} $, so that $ \vect{\theta}=[\vec \left( [ \matr{U}; \matr{V}  ] \right);\diag{\matr{\Sigma}}] $ and $\matr{E}$ is a noise matrix. If $ \tens{Y}' $ is a third order tensor, its Canonical Polyadic (CP) decomposition \cite{Comon2009}, \cite{Kolda2009} is written as $ \tens{Y}'= \left(\matr{A}',\matr{B}',\matr{C}'\right)+\tens{E}' $, meaning that
$\tenselem{Y'}_{ijk} = \sum_{r=1}^R A'_{ir} B'_{jr} C'_{kr}+\tenselem{E}_{ijk}'$, $ \vect{\theta}'=\vec\left( [\matr{A}' ; \matr{B}' ; \matr{C}'] \right)$, and again $ \tens{E}' $ is a noise array.

In the case where $ \vect{\theta} $ and  $ \vect{\theta}' $ are not coupled, they can be obtained separately from each array, generally non uniquely in the matrix case. On the other hand, if they are coupled, then the arrays can be processed jointly and parameters are then uniquely estimated, under mild additional assumptions \cite{Sorensen2015a}.\label{uniquenessPage}  
Here we consider that the coupling between $ \vect{\theta} $ and $ \vect{\theta}' $ is flexible: for example, we could have $ \matr{V} \approx \matr{B}$, or $ \matr{V}\approx \matr{W}\matr{B} $ for a known transformation matrix $ \matr{W} $. To formalize this we assume that the pair $ \vect{\theta} $, $ \vect{\theta}' $ is random and that we have at our disposal a joint probability density function (PDF) $ p(\vect{\theta},\vect{\theta}') $.

\paragraph*{Maximum \textit{a posteriori} estimator}since the pair $ \vect{\theta} $, $ \vect{\theta}' $ is random, we adopt a maximum \textit{a posteriori} (MAP) setting\footnote{We could also consider a minimum mean squared error setting but then we would need to evaluate $ p(\tens{Y},\tens{Y}') $, which is typically cumbersome.}. \label{pdfPage} In this setting, the approximation is defined as 
\begin{equation}
\begin{aligned}
\underset{\vect{\theta},\vect{\theta}'}{\arg\max}\; p(\vect{\theta},\vect{\theta}'\vert \tens{Y},\tens{Y}' )&=\underset{\vect{\theta},\vect{\theta}'}{\arg\max}\; p(\vect{\theta},\vect{\theta}', \tens{Y},\tens{Y}' ) \nonumber\\
&=\underset{\vect{\theta},\vect{\theta}'}{\arg\min}\;\Upsilon(\vect{\theta},\vect{\theta}')
\end{aligned}
\label{map}
\end{equation}
where $ \Upsilon(\vect{\theta},\vect{\theta}')=-\log p(\vect{\theta},\vect{\theta}', \tens{Y},\tens{Y}' ) $.  Conditioning on the parameters leads to a cost function that can be decomposed into a joint data likelihood term plus a term involving the coupling: 
 \begin{equation}
\Upsilon(\vect{\theta},\vect{\theta}')=-\log p(\tens{Y},\tens{Y}' \vert \vect{\theta},\vect{\theta}')-\log p(\vect{\theta},\vect{\theta}').
\label{gen_cost}
\end{equation}
Under the following simplifying hypotheses underlying the Bayesian approach, the  MAP estimator is given as the minimizer of a three-term cost function :
\begin{align}
\underset{\vect{\theta},\vect{\theta}'}{\arg\min}\;\Upsilon(\vect{\theta},\vect{\theta}')=\qquad\qquad\qquad\qquad\qquad\qquad\qquad\nonumber\\ 
\underset{\vect{\theta},\vect{\theta}'}{\arg\min}\;\left[ -\log p(\tens{Y}\vert \vect{\theta})-\log p(\tens{Y}'\vert \vect{\theta}')-\log p(\vect{\theta},\vect{\theta}')\right].
\label{simp_map}
\end{align}

\begin{description}
\item[H1] \emph{Conditional independence of the data}: the data arrays $ \tens{Y} $ and $ \tens{Y}' $ are independent of $ \vect{\theta}' $ and $ \vect{\theta} $ respectively, if they are conditioned on $ \vect{\theta} $ and $ \vect{\theta}' $. We suppose also that they are conditionally independent on each other. This results in the following $ p(\tens{Y}\vert \tens{Y}',\vect{\theta},\vect{\theta}')=p(\tens{Y}\vert \vect{\theta}) $ and $ p(\tens{Y}'\vert \tens{Y},\vect{\theta},\vect{\theta}')=p(\tens{Y}'\vert \vect{\theta}') $.
\item[H2] \emph{Independence of uncoupled parameters}: all parameters except the coupled parameters are independent. In the joint SVD and CP cases, this means that $p(\vect{\theta},\vect{\theta}')=p(\matr{V},\matr{B})p(\matr{U})p(\matr{\Sigma})p(\matr{A})p(\matr{C})$ if coupling  exists only in the second mode. 
\item[H3] \emph{On the priors}: trivially, the joint distribution of the coupled parameters, \textit{e.g.} $ p(\matr{V},\matr{B})  $ needs to be known or, at least, one of the conditional distributions, namely $ p(\matr{V}\vert\matr{B})  $ or  $ p(\matr{B}\vert\matr{V})$. The marginal priors on the uncoupled and on the conditioning parameters ($ p(\matr{U}),\,\ldots,\, p(\matr{C})$) are assumed either to be known or to be flat on some domain of definition.
\item[H4] \emph{Likelihoods}: the conditional probabilities (or likelihoods) $ p(\tens{Y}\vert \vect{\theta}) $ and $ p(\tens{Y}'\vert \vect{\theta}') $ are known, at least up to a scale parameter. In a MAP setting, this indirectly sets the weights which will be given to each data array in $ \Upsilon\left(\vect{\theta},\vect{\theta}'\right)  $.
\end{description}

Hypotheses H3 and H4 are assumed so that all terms in this cost function are defined. H2 is assumed so that the probabilistic dependence of the parameters defines the coupling between various  latent models.

The framework presented above contains as specific cases the following:
\begin{itemize}
\item \emph{Hybrid estimation:} if we consider that the conditional distribution $ p(\matr{V}\vert\matr{B}) $ is known and that $ \matr{B} $ is deterministic and defined in a domain $ \matr{\Omega}_{\matr{B}} $, then we have to minimize $ \Upsilon_{H}(\vect{\theta},\vect{\theta}')=-\log p(\tens{Y}\vert \vect{\theta})-\log p(\tens{Y}'\vert \vect{\theta}')-\log p(\matr{V},\matr{B}') $ subject to $ \matr{B}\in \matr{\Omega}_{\matr{B}} $, which is a hybrid (random/deterministic) coupled approximation problem. This is the framework implicitly considered in \cite{Seichepine2014}. Note that to enforce symmetry of roles of the coupled parameters, for example if $ \matr{V} $ and $ \matr{B} $ are noisy versions of each other, we can consider that they are similar versions of the same deterministic latent factor $ \matr{V}^{\star} $. This symmetric approach has the advantage that it can be trivially extended to more than two coupled models.
\item \emph{Coupled approximation:} the usual assumption that the coupled factors are equal, $ \matr{V}=\matr{B} $, can be obtained by setting a Dirac's delta prior $ p(\vect{\theta},\vect{\theta}')=\delta({B-V}) $. The MAP problem becomes a MLE problem with equality constraints
\begin{equation}
\begin{array}{cc}
\text{maximize} & \log p(\tens{Y};\vect{\theta})+\log p(\tens{Y}';\vect{\theta}') \\ 
\text{with respect to (w.r.t.)} &  \vect{\theta},\,\vect{\theta}' \\ 
\text{subject to}& \matr{V}=\matr{B}
\end{array} 
\label{joint_MLE}
\end{equation}
which corresponds to the coupled approximation framework with general cost functions. Different versions of this approach are proposed in \cite{Singh2008,Yilmaz2011,Ermics2012}. Under the assumption of Gaussian likelihoods, we have the standard coupled approximation framework \cite{Acar2011}. 
\item \emph{Exactly coupled decompositions:} additionally to exact coupling of one of the factors, if the data sets are not corrupted by noise, the likelihoods are also products of delta functions and we have to solve an exactly coupled decomposition problem \cite{Sorensen2015a}, \cite{Sorensen2015b}.
\end{itemize}

In the next section we give many examples of possible joint decomposition problems and their MAP or hybrid MAP/MLE objective functions. 

\section{Flexible coupling models}
\label{sec:examples}
In what follows we consider that the parametric models underlying the data arrays are two CP models $ \tens{X}=\left(\matr{A},\matr{B},\matr{C}\right)  $ and $ \tens{X}'=\left(\matr{A}',\matr{B}',\matr{C}'\right)  $ with dimensions $ I $, $ J $, $ K $ and $ I' $, $ J' $, $ K' $ and number of components (\textit{i.e.} number of matrix columns) $ R $ and $ R' $ respectively. We consider that the coupling occurs between matrices $ \matr{C} $ and $ \matr{C'} $. We illustrate this framework with three different examples: general joint Gaussian, hybrid Gaussian and non Gaussian models for the parameters.

\subsection{Joint Gaussian modeling}\label{jointGaussModel-sec}
A general joint Gaussian model comprising coupled and uncoupled variables is given by the following expression:
\begin{equation}
\matr{M}\left[\vect{\theta} ; \vect{\theta}'\right]=\matr{\Sigma}\vect{u}+\vect{\mu}
\label{joint_gauss}
\end{equation}
where $ \matr{M} $ is a matrix defining the structural relations between variables, $ \vect{u} $ is a white Gaussian vector with zero mean and unit variances, $ \matr{\Sigma} $ is a diagonal matrix of standard deviations and $ \vect{\mu} $ is a constant vector. A condition for the pair $ (\vect{\theta},\vect{\theta}') $ to define a joint Gaussian vector is the left invertibility of $\matr{M}$. Under this condition we have $ \left[\vect{\theta}; \vect{\theta}'\right]\sim \tens{N} \lbrace {}^{\dagger}\!\matr{M}\vect{\mu}, \matr{R}\rbrace  $, where $ \matr{R}=(\leftinv{M})\matr{\Sigma}\matr{\Sigma}(\leftinv{M}^\T)  $ is the covariance matrix of the joint vector.

Considering that the CP models $ \tens{X} $ and $ \tens{X}' $ are measured with zero mean independent additive Gaussian noise, with variances $ \sigma_{n}^{2} $ and $ \sigma_{n}'^{2} $, respectively,  we have the following MAP objective function:
\begin{equation}
\begin{aligned}
&\Upsilon\left(\vect{\theta},\vect{\theta}'\right)=\\
&\qquad\frac{1}{\sigma_{n}^{2}}\|\tens{Y}-(\matr{A},\matr{B},\matr{C}) \|_{F}^{2}+\frac{1}{\sigma_{n}'^{2}}\|\tens{Y}'-(\matr{A}',\matr{B}',\matr{C}') \|_{F}^{2}\\
&\qquad+ \| [ \vect{\theta}; \vect{\theta}']  -{}^{\dagger}\!\matr{M}\vect{\mu}\|_{\matr{R}}^{2}
\end{aligned}
\label{map_j_gauss}
\end{equation}
where $ \|\cdot\|_{F} $ and $ \|\cdot\|_{\matr{R}} $ are respectively the Frobenius norm and $\matr{R}$-weighted Frobenius norm.
We give below two examples of applications of this approach.
\paragraph*{Shared components}a usual problem in multimodal data fusion is that some components are not present in all modalities, thus we have some components which are shared and some which are specific to each modality. Suppose $\vect{\theta}=[\vect{\theta_u}; \vect{c}_1]$ and $\vect{\theta}'=[\vect{\theta_u'}; \vect{c}_1']$, where $\vect{c}_1$ and $\vect{c}_1'$ are coupled, whereas $\vect{\theta_u}$ and $\vect{\theta_u}'$ are not.

Supposing zero mean marginal Gaussian priors, we may write \vspace{-1ex}
\begin{eqnarray}\label{uncoupled-eq}
\vect{\theta}_u = \matr{\Sigma}_u \vect{u}, \quad \vect{\theta}_u' = \matr{\Sigma}_u' \vect{u}'\\
\vect{c}_1 - \vect{c}_1' = \matr{\Sigma}_c \vect{u}_1, \quad \vect{c}_1' = \matr{\Sigma}_c' \vect{u}_1'
\label{coupled-eq}
\end{eqnarray}
where $\vect{u}$, $\vect{u}'$, $\vect{u}_1$ and $\vect{u}_1'$ are zero-mean unit variance independent Gaussian variables, and where $\matr{\Sigma}_u$, $\matr{\Sigma}_u'$, $\matr{\Sigma}_c$ and $\matr{\Sigma}_c'$ define corresponding covariance matrices.
With an obvious definition of $\matr{M}$ and $\matr{\Sigma}$, equations (\ref{uncoupled-eq}-\ref{coupled-eq}) can be grouped in a unique block equation of the form:

\small
$$
\matr{M} \left[\begin{array}{cccc} \vect{\theta}_u\\ \vect{c}_1\\ \vect{\theta}_u'\\  \vect{c}_1'\end{array}\right] = \matr{\Sigma} \left[\begin{array}{cccc} \vect{u}\\  \vect{u}_1\\ \vect{u}'\\  \vect{u}_1'\end{array}\right].
$$
\normalsize

\paragraph*{Dynamical models} in some cases, more than two data arrays are present and, as a consequence, there is a large number of degrees of freedom in the definition of the couplings. If the data arrays are measured in time, then a natural coupling can be defined through a dynamical model on the factors \cite{Rogers2013}. Finding the MAP estimator in this case amounts to solve a smoothing problem, since the temporal relations embedded in the dynamical model will impose some smoothness on the estimated factors.

\subsection{Hybrid Gaussian coupling}\label{hybridGaussCoupling-sec}

When no prior information is given on some parameters, full joint Gaussian modeling is inadequate, since we cannot define means and variances of the model. Instead, we consider these parameters as deterministic, while the other parameters are still stochastic with Gaussian priors. We call this model \emph{hybrid Gaussian}. This is the main focus of this paper since most applications will fall in this category of coupling. Indeed, it covers the scenario where only one factor, say $\matr{C}$, is coupled to another, say $\matr{C}'$, with transformation matrices and independent and identically distributed (i.i.d.) Gaussian additive noise:
\begin{equation}
\matr{H} \matr{C}=\matr{H}' \matr{C}' + \vect{\Gamma}, \quad \Gamma_{ij} \sim\mathcal{N}(0,\sigma_c^2)
\label{hybridgaussian}
\end{equation}
for some transformation matrices $\matr{H}$ and $\matr{H}'$. If relation (\ref{hybridgaussian}) is the only known relationship between parameters, it means that only the conditional probability $p(\matr{C}|\matr{C}')$ is known, and that $\matr{C}'$ is either deterministic or has a flat non-informative prior. 

This coupling model cannot be written with zero noise variance if the transformation matrices are tall (\textit{i.e.} more rows than columns),  as the set of feasible parameters would be reduced to the trivial set. This is important because using zero noise variance as a working hypothesis as in \cite{Acar2013a} leads to a bias in the hybrid Gaussian coupling estimation setting when transformation matrices are tall.

Even more critical is the norm of the coupled factors. Without proper normalization, the CP decomposition is invariant w.r.t. the norm of columns of $\matr{C}$ and $\matr{C}'$ as it can be incorporated in the other factors. This means that the variance of $\matr{\Gamma}$ is not well defined when defining the joint tensor decomposition problem with (\ref{hybridgaussian}). It is necessary to add the information on the norm of columns of the coupled factors in the hybrid model (\ref{hybridgaussian}), or to consider the coupling variance as an unknown parameter. This will also impact algorithms designed in Section \ref{sec:alg}.

\paragraph*{Sampling a continuous function}as an example, an important problem in multimodal data fusion is related to sampling. Different measurement devices have different sampling frequencies or even different non uniform sampling grids. In some situations, the continuous functions being measured can be approximated by a common function $ c(t) $. For two sampled vectors $ \vect{c} $ and $ \vect{c}' $, their relation with the continuous function can be obtained with an interpolation kernel (see the general description in \cite{Aldroubi2001})
\begin{equation}
c(t)\approx\sum\limits_{k=1}^{K} c_{k} h(t,t_{k})\approx\sum\limits_{k'=1}^{K'} c_{k}' h'(t,t_{k'})
\label{cont_func}
\end{equation}
for some kernels $ h(\cdot,\cdot) $ and $ h'(\cdot,\cdot) $ and for sampling times $ \lbrace t_{k} , k\in  1,\,\cdots,\,K  \rbrace$ and $ \lbrace t_{k'} , k'\in  1,\,\cdots,\,K'  \rbrace$. Therefore, we can impose a new common sampling grid of size $ L $ where both interpolations should match. This leads to the linear relations $\matr{H} \vect{c}\approx\matr{H}' \vect{c}'$, where $ H_{lk}=h(t_{l},t_{k}) $, $ H'_{lk'}=h(t_{l},t_{k'}) $ and $ \lbrace t_{l} , l\in  1,\,\cdots,\,L  \rbrace $. For example, in the coupled CP models, when $ \matr{C} $ and $ \matr{C}' $ have different dimensions due to different sampling rates, their coupling is governed by:

\begin{equation}
\matr{H} \vect{c}=\matr{H}' \vect{c}' + \vect{\gamma}.
\label{oversamp}
\end{equation} 

\subsection{Non Gaussian conditional coupling}\label{NonGaussCondCoupling-sec}
Non trivial couplings expressing similarity between the factors $ \matr{C} $ and $ \matr{C}' $ can be addressed by considering that $ \matr{C}' $ is deterministic and that the coupling is given by a conditional non Gaussian distribution, $p\left(\matr{C}\vert \matr{C}' \right)$.

\paragraph*{Impulsive additive coupling}as a first example, we can assume that each element in $ \matr{C} $ is a version of $ \matr{C}' $ corrupted by i.i.d. impulsive noise:
\begin{equation}
C_{ij}=C_{ij}'+\Gamma_{ij}
\label{imp_noise}
\end{equation}
where $ \Gamma_{ij} $ follows a Laplacian distribution $ p(\Gamma_{ij})=(1/2\delta_{c})\exp(-\vert\Gamma_{ij}\vert/\delta_{c}) $ with scale parameter $ \delta_{c} $ or a Cauchy distribution $ p(\Gamma_{ij})=1/\lbrace\pi\delta_{c}[1+(W_{ij}/\delta_{c})^{2}]\rbrace $. The objective functions to be minimized are then 
\begin{equation}
\begin{aligned}
\Upsilon\left(\vect{\theta},\vect{\theta}'\right)=&-\log p(\tens{Y}\vert \vect{\theta})-\log p(\tens{Y}'\vert \vect{\theta}')+\\
&\begin{cases}
+(1/2\delta_{c})\| \matr{C}-\matr{C}'\|_{1}&\text{(Laplace)}\\
-\sum\limits_{ij}^{IJ}\log \lbrace 1+[(
C_{ij}-C'_{ij})/\delta_{c}]^{2}\rbrace &\text{(Cauchy)}
\end{cases}
\end{aligned}
\label{map_imp}
\end{equation}
where $ \| \cdot\|_{1} $ is the $\ell_{1}$ norm. The first penalty was considered in \cite{Seichepine2014} in a collective matrix factorization context. Both cost functions imply a small number of large discrepancies between $ \matr{C} $ and $ \matr{C}' $.

\paragraph*{Positive general coupling} when $ C_{ij}>0 $ and $ C'_{ij}>0 $, an additive noise in the coupling may not be the best option, since, to ensure $ C_{ij} $ are positive, the support of the additive term has to depend on the values of $ C'_{ij} $, which is not a realistic assumption on a perturbation term. Therefore, other alternatives naturally ensuring positiveness can be considered. For example, the Tweedie's distribution \cite{Jorgensen1992} which contains Poisson, Gamma and inverse-Gaussian distributions as special cases (the Gaussian distribution is a limiting special case). In general, the PDF of the Tweedie's distribution has no analytical form, thus we cannot directly use it to write down a coupling term in the MAP objective function. However, if we consider that the coupling between $ C_{ij}>0 $ and $ C'_{ij}>0 $ is strong (dispersion $ \phi_{c} $ is small), then a saddle point approximation can be used \cite{Jorgensen1992}: 
\begin{equation}
p(C_{ij}\vert C'_{ij})\approx (2\pi\phi_{c}C_{ij}^{\beta_{c}})^{-1/2}\exp[-d_{\beta_{c}}(C_{ij}\vert C_{ij}')/\phi_{c}] 
\label{tweedie_saddle}
\end{equation}
where $ \beta_{c} $ is a shape parameter ($ \beta_{c}=1,2,3 $ for the Poisson, Gamma and inverse-Gaussian distributions respectively) and $ d_{\beta_{c}} $ is the beta divergence \cite{Yilmaz2012}: 
\begin{equation}
\begin{aligned}
d_{\beta_{c}}(C_{ij}\vert C_{ij}')=\qquad\qquad\qquad\qquad\qquad\qquad\qquad\qquad\qquad\\
\frac{C_{ij}'^{1-\beta_c}}{ \kappa(\beta_c)}\left[C_{ij}^{2-\beta_c}C_{ij}'^{\beta_c-1}
-C_{ij}(2-\beta_c)+C'_{ij}(1-\beta_c)\right]
\end{aligned}
\label{tweedie}
\end{equation}
where $ \kappa(\beta)=(1-\beta)(2-\beta) $. Under this conditional distribution, assuming that the CP model is positive ($ \matr{A} $, $ \matr{B} $, $ \matr{A}' $ and $ \matr{B}' $ are also positive) and that the data distributions are also of Tweedie type with shapes $ \beta $ and $ \beta' $ and dispersions $ \phi $ and $ \phi' $, we have the following hybrid MAP/MLE objective: \vspace{-1ex plus 0.5ex}
\begin{equation}\small
\begin{aligned}
\Upsilon(\vect{\theta},\vect{\theta}')=\qquad\qquad\qquad\qquad\qquad\qquad\qquad\qquad\qquad\qquad\\ 
\frac{1}{\phi}\left[\sum\limits_{ijk}^{IJK} d_{\beta}(\tenselem{Y}_{ijk}\vert\tenselem{X}_{ijk}) \right]+\frac{1}{\phi'}\left[\sum\limits_{ijk}^{I'J'K'} d_{\beta'}(\tenselem{Y}_{ijk}\vert\tenselem{X}_{ijk}) \right]\\ 
+\sum\limits_{ij}^{KR}\left[(\beta/2)\log\left(C_{ij}\right)+(1/\phi_{c})d_{\beta}(C_{ij}\vert C_{ij}')\right].
\end{aligned}
\label{MAP_tweedie}
\end{equation}

\section{Algorithms for coupled tensor decompositions with flexible couplings}
\label{sec:alg}
 
To fix the ideas, our discussion will be limited to coupled CP models where the coupling occurs in the third mode, that is, between the third factor matrices. An algorithm for the joint Gaussian coupling revolving around vectorization of factors matrices -- the coupling itself being applied to  vector $\vect{\theta}$ -- has been  derived, but it is bound to be computationally heavy since it requires the solution of large linear systems of equations at each iteration. Thus, it will not be presented in this work. An algorithm for the $\ell_{1}$ coupling model has already been presented in \cite{Seichepine2014}, and its counterpart for the Cauchy distribution can be obtained with a gradient approach, since the coupling prior distribution is smooth. We will refrain from detailing these algorithms. In this section we focus on algorithms to minimize the objective functions for the remaining presented models, namely, hybrid Gaussian models with i.i.d. Gaussian likelihoods (\ref{hybridgaussian}) and the Tweedie's model (\ref{MAP_tweedie}).
 
\subsection{Hybrid Gaussian modeling and alternating least squares (ALS)}\label{hybridGaussModelALS-sec}
Under hybrid Gaussian coupling assumption, the MAP estimator is the minimizer of the following cost function\footnote{For numerical reasons, if $\sigma_c$ is much smaller than $\sigma_n$ or $\sigma_n'$, it will be preferred in practice to set $\matr{H}\matr{C}=\matr{H}'\matr{C}'$.}:
\begin{equation}
\begin{aligned}
&\quad\Upsilon\left(\vect{\theta},\vect{\theta}'\right)=\frac{1}{\sigma_{n}^{2}}\|\tens{Y}-(\matr{A},\matr{B},\matr{C}) \|_{F}^{2}~+\\
&\frac{1}{\sigma_{n}'^{2}}\|\tens{Y}'-(\matr{A}',\matr{B}',\matr{C}') \|_{F}^{2} + \frac{1}{\sigma_c^2}\|\matr{H}\matr{C}-\matr{H}'\matr{C}'\|_{F}^2.
\end{aligned}
\label{cost_hybrid}
\end{equation}
To minimize (\ref{cost_hybrid}), it is possible to resort to standard algorithms matched to convex optimization, which generally requires to tune additional parameters like stepsize. Here we chose to resort to a modified version of the widely used  ALS \cite{Comon2009}, which is simple to implement.
\smallskip

\paragraph*{Joint update of  coupled factors}to build the ALS algorithm, the gradient of $\Upsilon$ w.r.t. each factor matrix needs to be computed, and set to zero. For uncoupled factors, this leads to the standard uncoupled ALS update rules, see Algorithm \ref{als_alg}.

Now for the coupled factors $\matr{C}$ and $\matr{C}'$, a simultaneous update ensuring symmetry in the treatment of both data sets requires to solve the following system: \vspace{-1ex plus 0.5ex}
\begin{equation}\small
\left\{
\begin{array} {lcl}
\frac{1}{\sigma_c^2}\matr{H}^\T\matr{H}\matr{C} + \frac{1}{\sigma_n^2}\matr{C}\matr{D} - \frac{1}{\sigma_c^2}\matr{H}^\T\matr{H}'\matr{C}' & = & \frac{1}{\sigma_n^2}\matr{Y_{(3)}}\matr{D} \\
\frac{1}{\sigma_c^2}\matr{H}'^\T\matr{H}'\matr{C}' + \frac{1}{\sigma_{n}'^2}\matr{C}'\matr{D}' - \frac{1}{\sigma_c^2}\matr{H}'^\T\matr{H}'\matr{C} & = & \frac{1}{\sigma_{n}'^2}\matr{Y_{(3)}}'\matr{D}'  
\end{array}
\right.
\label{eqn_gaussian}
\end{equation}\normalsize
where  $\matr{D} \!= \!\left(\matr{B}\khatri\matr{A}\right)^\T \! \left(\matr{B}\khatri\matr{A}\right)$, and $\matr{D}' \!=\! \left(\matr{B}'\khatri\matr{A}'\right)^\T \! \left(\matr{B}'\khatri\matr{A}'\right)$.
This system is costly to solve for $\{\matr{C}, \matr{C}'\}$. It contains indeed two coupled Sylvester equations. A straightforward approach teaches us to vectorize parameter matrices $\matr{C}$ and $\matr{C}'$ and to solve the following stacked large linear system of equations w.r.t. $\{\matr{C}, \matr{C}'\}$: 
\begin{equation}
\matr{G}\,\vec([\matr{C};\matr{C}'])=\vec\left(\left[\frac{1}{\sigma_n^2}\matr{Y_{(3)}}\matr{D};\frac{1}{\sigma_{n}'^2}\matr{Y_{(3)}}'\matr{D}'\right]\right)
\label{full_als}
\end{equation}
 where $\matr{G}=[\matr{G}_{11}, \matr{G}_{12} ; \matr{G}_{21},  \matr{G}_{22}]$, with
 
\small
\begin{equation}
\begin{array}{cl}
\matr{G}_{11} = &\frac{1}{\sigma_c^2}\matr{I}_R\kron\matr{H}^\T\matr{H} + \frac{1}{\sigma_n^2}\matr{D}^\T\kron\matr{I}_K \\
\matr{G}_{22} = & \frac{1}{\sigma_c^2}\matr{I}_R\kron\matr{H}'^\T\matr{H}' + \frac{1}{\sigma_{n}'^2}\matr{D}'^\T\kron\matr{I}_K \\
\matr{G}_{12} = & - \frac{1}{\sigma_c^2} \matr{I}_R \kron \matr{H}^\T\matr{H}' \\
\matr{G}_{21} = &- \frac{1}{\sigma_c^2}  \matr{I}_R\kron \matr{H}'^\T\matr{H}.
\end{array}
\end{equation}\normalsize

\textbf{Remark.}~ Note that we present here an alternating algorithm for one example of a hybrid Gaussian coupling. Another possible type of coupling is when the transformation matrices act on the right of the factors, \textit{i.e.} when the transformation is in the component space. Then the previous algorithm should be modified because full vectorization is not needed, as the Sylvester structure of the two coupled equations becomes a simple stacked linear system.

\smallskip

\paragraph*{Full alternating solution} a full alternating solution can be obtained by solving sequentially the two equations in (\ref{eqn_gaussian}). At the $ k $-th iterate we obtain the estimate $ \hat{\matr{C}}_{k} $ by solving

\small
\begin{equation}
\frac{1}{\sigma_c^2}\matr{H}^\T\matr{H}\matr{C} + \frac{1}{\sigma_n^2}\matr{C}\matr{D} = \frac{1}{\sigma_c^2}\matr{H}^\T\matr{H}'\hat{\matr{C}}'_{k-1} + \frac{1}{\sigma_n^2}\matr{Y_{(3)}}\matr{D}
\label{eqn_gaussian_1}
\end{equation}\normalsize
 w.r.t. $ \matr{C} $, where $\hat{\matr{C}}'_{k-1}  $ is the previous estimate, considered as a fixed constant matrix for $ \matr{C}' $. The estimate $ \hat{\matr{C}}_{k}' $ is then obtained by solving
\begin{equation}\textstyle
\frac{1}{\sigma_c^2}\matr{H}'^\T\matr{H}'\matr{C}' + \frac{1}{\sigma_{n}'^2}\matr{C}'\matr{D}' = \frac{1}{\sigma_c^2}\matr{H}'^\T\matr{H}'\hat{\matr{C}}_{k} + \frac{1}{\sigma_{n}'^2}\matr{Y_{(3)}}'\matr{D}'
\label{eqn_gaussian_2}
\end{equation}
w.r.t. $ \matr{C}' $. This approach leads to a much faster computation of each iteration compared to the simultaneous update previously described, since any algorithm efficiently solving the Sylvester equation can be used  \cite{DattS91:laa,GoluNV79:tac}. 

\smallskip

\paragraph*{Scale ambiguity and normalization} since CP decomposition does not directly impose scaling on the estimated factors, the coupling term in the MAP objective function can be decreased by rescaling the  coupled factors. Therefore, without proper normalization on the factors updates, the norms of the columns of the coupled factors $\matr{C}$ and $\matr{C}'$ may tend to zero. In addition, if the norms of the columns of $\matr{C}$ and $\matr{C}'$ tend to zero, the coupling term will become negligible when compared to the likelihood terms. As a consequence, the updates will tend to the decoupled solution. Moreover, $\sigma_c^2$ will not represent the actual variance in the coupling model, and a scaling matrix should be introduced in the covariance matrix of $\matr{\Gamma}$.

A first solution to overcome this problem would be to normalize the columns of $\matr{C}$ directly in the designed algorithms, for example using the $\ell_{1}$ or $\ell_{2}$ norms. But this means adding a norm constraint on the coupled factor which is already heavily constrained by the coupling relationship. We thus made the choice to normalize $\matr{A}$ and $\matr{B}$ to indirectly impose some normalization of $\matr{C}$, and only normalize $\matr{A}'$ since $\matr{C}'$ will be determined partially by the coupling model.

\smallskip

\paragraph*{Initialization and stopping}the MAP objective function for the coupled CP model is non convex, as a consequence, multiple local maxima may be present. To obtain an estimate close to the MAP estimator (the global maximum) a standard approach is to carry out ALS multiple times with different initializations, and then pick the solution giving the best MAP objective value \cite[pp.62-63, p.122]{Bro1998}. There are two options for initializing the coupled algorithms, namely starting directly from random guesses (cold start), or refining those guesses in two uncoupled ALS (warm start). We found that these two approaches are not equivalent, and starting from the random guesses may lead to larger bias and slower decrease of the cost function when the coupling is weak (see Appendix \ref{app_3}). However, before initializing the coupled ALS procedure with the uncoupled results, it is necessary to correct the permutation ambiguity which are intrinsic to the CP model. To achieve this, one of the estimated CP models is fixed and the components of the other CP model are permuted such that the value of the coupling term $ \|\matr{H}\matr{C}-\matr{H}'\matr{C}'\|_{F}^2 $ is minimized. This can be done by solving an assignment problem, which in practice is carried out in polynomial time with the Hungarian algorithm \cite{Munkres1957}.

At each run, the algorithm stops at iterate $ k $ if either the relative decrease $ \Delta\Upsilon_{k}=\left\vert\frac{\Upsilon_{k}-\Upsilon_{k-1}}{\Upsilon_{0}}\right\vert $ achieves $ \Delta\Upsilon_{\min} $ or a maximum number of iterates $ k_{\max} $ has been reached.

\begin{algorithm}
\caption{ALS algorithm for the hybrid Gaussian coupled model}
\begin{algorithmic}[1]
\REQUIRE {$\tens{Y}$, $\tens{Y}'$, $\sigma_c$, $\sigma_n$, $\sigma_n'$.}
\STATE Apply two uncoupled ALS with random initializations to obtain $ \hat{\matr{A}}_{0} $, $ \hat{\matr{B}}_{0} $, $ \hat{\matr{C}}_{0} $, $ \hat{\matr{A}}_{0}' $, $ \hat{\matr{B}}_{0}' $ and $ \hat{\matr{C}}_{0}' $.
\STATE Permute initialization factors columns to have minimal $ \|\matr{H}\hat{\matr{C}}_{0}-\matr{H}'\hat{\matr{C}}_{0}'\|_{F}^2 $.
\STATE Evaluate the initial cost function $ \Upsilon_{0} $ (\ref{cost_hybrid}). 
\STATE Set $ k=0 $, $ \Delta\Upsilon=+\infty $.
\WHILE{$ k<k_{\mathrm{max}} $ and $ \Delta\Upsilon_{k}>\Delta\Upsilon_{\text{min}} $,} 
\STATE{k:=k+1.}
\STATE Update $\hat{\matr{A}}_{k}$: \\
$\hat{\matr{A}}_{k} = \matr{Y}_{(1)}\,\left[ \left(\hat{\matr{C}}_{k-1}\khatri\hat{\matr{B}}_{k-1}\right)^{\T}\right]^{\dagger}$. \\
Normalize columns of $\hat{\matr{A}}_{k}$.
\STATE Similarly update $\hat{\matr{A}}_{k}'$, $\hat{\matr{B}}_{k}$ and $\hat{\matr{B}}_{k}'$.\\ Normalize $\hat{\matr{A}}_{k}'$ and $\hat{\matr{B}}_{k}$, $\hat{\matr{B}}_{k}'$ remaining unnormalized.
\STATE Solve (\ref{full_als}) to obtain the joint update $ [\hat{\matr{C}}_{k}; \hat{\matr{C}}_{k}']  $ or (\ref{eqn_gaussian_1}) and (\ref{eqn_gaussian_2}) sequentially.
\STATE Compute cost function $\Upsilon_{k}$ (\ref{cost_hybrid}) and its relative $\Delta\Upsilon_{k}$.
\ENDWHILE
\RETURN $ \hat{\matr{A}}_{k} $, $ \hat{\matr{B}}_{k} $, $ \hat{\matr{C}}_{k} $, $ \hat{\matr{A}}_{k}' $, $ \hat{\matr{B}}_{k}' $ and $ \hat{\matr{C}}_{k}' $.
\end{algorithmic}
\label{als_alg}
\end{algorithm}

\subsection{Coupled decompositions of large data sets}\label{coupledDecLarge-sec}
Let us consider again the hybrid Gaussian problem (\ref{hybridgaussian}) with the simple coupling
\begin{equation}
\matr{C}=\matr{C}'+\matr{\Gamma}
\label{simple_coupling}
\end{equation}
where $ \matr{\Gamma} $ is an i.i.d. Gaussian matrix with variance of each element $ \sigma_{c}^{2} $ and $\matr{C}'$ has columns of given norm. A natural question that raises from previous discussions is scalability of coupled decompositions. If we disregard the coupling of the tensors, a common strategy to retrieve the CP models in a more computationally efficient way is to compress the data arrays, decompose the compressed tensors, and then uncompress the obtained factors \cite[p. 92]{Bro1998}.

\smallskip

\paragraph*{Uncoupled compression} to compress data arrays, the higher order singular value decomposition (HOSVD)  \cite{DeLathawer2000} can be used. The HOSVD is given by three orthogonal factors $ \matr{U} $, $ \matr{V} $, $ \matr{W} $ and a core tensor $ \tens{G} $ with dimensions $ R_{1}\times R_{2}\times R_{3}  $. 
If we assume that data are noiseless, we have
\begin{equation}
\begin{aligned}
\tenselem{X}_{ijk}&=\sum\limits_{rst}^{R_{1}R_{2}R_{3}} U_{ir} V_{js} W_{kt}\tenselem{G}_{rst} \\ \tenselem{X}_{ijk}'&=\sum\limits_{rst}^{R_{1}'R_{2}'R_{3}'}U_{ir}' V_{js}' W_{kt}'\tenselem{G}_{rst}'
\label{hosvd_1}
\end{aligned}
\end{equation}
which we can denote as $ \tens{X}=(\matr{U},\matr{V},\matr{W})\tens{G} $ and $ \tens{X}'=(\matr{U}',\matr{V}',\matr{W}')\tens{G}'$. The noiseless tensor models $ \tens{X} $ and $ \tens{X}' $ can be seen as array representations of multilinear operators, thus the HOSVD  described above corresponds to changes of coordinates respectively from $\tens{X}$ and $ \tens{X}' $ to $\tens{G}$ and $\tens{G}'$. If the multilinear subspaces spanned by each mode's vectorized slices of the original tensors are of low dimension, then the original tensors lie in the span of a small number of basis vectors in each mode, \textit{i.e.} $ (I,J,K)>(R_{1},R_{2},R_{3}) $ and  $ (I',J',K')>(R_{1}',R_{2}',R_{3}') $. The core tensors $ \tens{G} $ and $ \tens{G}' $ are rotations of $ \tens{X} $ and $ \tens{X}' $ with many zeros after a certain index in each mode. 

This approach is valid with equalities in (\ref{hosvd_1}), if tensors $\tens{X}$ and $\tens{X}'$ are not corrupted by noise. If additive measurement noise is present but its variance is not too large, HOSVD orthogonal factors explaining most (but not all) of the variance in tensor data can be accurately obtained from the data arrays $\tens{Y}$ and $\tens{Y}'$ by retaining only the first columns of the left singular vectors of the unfolded data arrays 
\begin{equation}
\left\{
\begin{aligned}
\matr{U}^{\mathrm{SVD}} \matr{N_1} = &\text{ SVD}\left[ \matr{Y}_{(1)} \right]& \\
\matr{V}^{\mathrm{SVD}} \matr{N_2} = &\text{ SVD}\left[ \matr{Y}_{(2)} \right]& \\
\matr{W}^{\mathrm{SVD}} \matr{N_3} = &\text{ SVD}\left[ \matr{Y}_{(3)} \right]
\label{hosvd_2}
\end{aligned}
\right. .
\end{equation}
The estimated factor matrices are $ \matr{U}=\matr{U}_{1:R_{1}}^{\mathrm{SVD}} $, $ \matr{V}=\matr{V}_{1:R_{2}}^{\mathrm{SVD}} $ and $ \matr{W}=\matr{W}_{1:R_{3}}^{\mathrm{SVD}} $. The core is estimated as 
\begin{equation}
\tenselem{G}_{rst} = \sum\limits_{ijk}^{IJK} \tenselem{Y}_{ijk}U_{ir}V_{js}W_{kt}.
\end{equation}
Now $\tens{G}$  is a compressed version of the  noisy tensor $\tens{Y}$.

We can now decompose the core tensors to obtain two CP models $ (\matr{A}_{c},\matr{B}_{c},\matr{C}_{c}) $ and $ (\matr{A}_{c}',\matr{B}_{c}',\matr{C}_{c}') $. The cores have much smaller dimensions than the original tensors, therefore obtaining the compressed CP models is much less computationally expensive. To retrieve the uncompressed CP models, we can use the fact that $ (\matr{U},\matr{V},\matr{W})\tens{G}=(\matr{U}\matr{A}_{c},\matr{V}\matr{B}_{c},\matr{W}\matr{C}_{c}) $, which means that the original factors can be recovered by inexpensive matrix multiplications
 \begin{equation}
\begin{aligned}
\matr{A}=\matr{U}\matr{A}_{c}\qquad\matr{B}=\matr{V}\matr{B}_{c}\qquad\matr{C}=\matr{W}\matr{C}_{c}.
\end{aligned}
\end{equation}
Note that the bottleneck with this approach is the computation of the three SVDs. This can be tackled efficiently using the Lanczos algorithms \cite{CullW85} or one of its variations\footnote{We adopted Algorithm 5.1 of \cite{Halko2011} with oversampling parameter of $2$ and one power iterate.} recently proposed in \cite{Halko2011}.

\paragraph*{Coupling in the compressed domain} if we apply the HOSVD to the noiseless tensor models $ \tens{X} $ and $ \tens{X}' $, we can express the coupling in the compressed space as
\begin{equation}
\matr{W}\matr{C}_{c}=\matr{W}'\matr{C}_{c}' +\matr{\Gamma}
\label{uncompressed_coupling}
\end{equation}
that is, we suppose $\matr{C}$ lies in the column space of $\matr{W}$ and $\matr{C}'$ in the column space of $\matr{W}'$. Therefore, we could assume the coupled approximation setting in the compressed space with the flexible coupling model (\ref{uncompressed_coupling}) above. However, if one of the data arrays is noisy, for example $ \tens{Y} $, its corresponding estimated HOSVD factor $\matr{W}$ may have large estimation errors and the coupling relation will be no longer true since $\matr{C}$ will not lie in the span of $\matr{W}$. On the other hand, if $ \sigma_{c}^{2} $ is small, factors $ \matr{C} $ and $ \matr{C}' $ are very similar, and as a consequence, the left singular vectors of the unfoldings $ \matr{X}_{(3)} $ and $ \matr{X}_{(3)}' $ are also similar. This indicates that we can obtain a joint orthonormal factor $ \matr{W}^{\!j}$ as left singular vectors of the stacked and weighted unfoldings
\begin{equation}
\matr{W}^{\!j} \, \matr{N_3} = \text{SVD}\left[ \frac{\matr{Y}_{(3)}}{\sigma_n}, \,\frac{\matr{Y}_{(3)}'}{\sigma_n'}\right].
\label{j_svd}
\end{equation} 
Now the uncompressed coupling relation (\ref{simple_coupling}) can be rewritten as $[\matr{W}^{\!j}]^{\T}\matr{C}=[\matr{W}^{\!j}]^{\T}\matr{C}' +[\matr{W}^{\!j}]^{\T}\matr{\Gamma}$, and is valid even in the presence of observation noise since $\matr{C}$ and $\matr{C'}$ do not need to belong to the column space of the jointly estimated factor $\matr{W}^{\!j}$.  Moreover, since compressed parameters are now defined in the noisy case as $\matr{C}_c:=[\matr{W}^{\!j}]^{\T}\matr{C}$  and  $\matr{C}'_c:=[\matr{W}^{\!j}]^{\T}\matr{C}'$, the similarity coupling appears directly in the compressed dimension
\begin{equation}
\matr{C}_{c}=\matr{C}_{c}' +\matr{\Gamma}_{c}
\label{compressed_coupling}
\end{equation} 
where $\matr{\Gamma}_{c} =[\matr{W}^{\!j}]^{\T} \,\matr{\Gamma}$ is a matrix of  same dimensions as  $\matr{C}_{c}$ and with Gaussian i.i.d. entries of variance $ \sigma_{c}^{2} $. The hybrid objective function in the compressed domain becomes

\small
\begin{equation}
\begin{aligned}
\Upsilon=\frac{1}{\sigma_{n}^{2}}\left\| \tens{G}- \left( \matr{A}_{c}, \matr{B}_{c}, \matr{C}_{c}\right)  \right\|_{F}^{2} +\frac{1}{\sigma_{n}'^{2}}\left\| \tens{G}'- \left( \matr{A}_{c}', \matr{B}_{c}', \matr{C}_{c}'\right) \right\|_{F}^{2}\\
\qquad\qquad +\frac{1}{\sigma_{c}^{2}} \left\| \matr{C}_{c}-\matr{C}_{c}' \right\|_{F}^{2}
\end{aligned}
\label{MAP_compressed}
\end{equation}\normalsize
which can be solved using the ALS algorithm (Algorithm \ref{als_alg}).

\subsection{Multiplicative updates for non Gaussian coupling of positive variables}\label{MultiUpdatesNonGaussian-sec}
To minimize the objective function (\ref{MAP_tweedie}) under the constraint that all estimates are nonnegative,  we can use a multiplicative update (MU) algorithm \cite{Lee2001}, which is a popular method in nonnegative matrix factorization. The idea of MU is to alternatively update the factors using a gradient descent algorithm where the stepsize is chosen as a function of the negative and positive parts of the gradients. As a consequence, the gradient descent update becomes a simple nonnegative multiplicative update \cite{Fevotte2011}.  For example, the update $ \hat{\matr{A}}_{k} $ at iteration $ k $ of the factor $ \matr{A} $ is
\begin{equation}
\hat{\matr{A}}_{k}=\hat{\matr{A}}_{k-1}\boxdot \left(\nabla_{\matr{A}}\Upsilon_k^{-} \boxslash \nabla_{\matr{A}}\Upsilon_k^{+} \right)
\label{mu_update_1}
\end{equation}
where $ \nabla_{\matr{A}}\Upsilon^{+} $ and $ \nabla_{\matr{A}}\Upsilon^{-} $ are the positive and negative parts of the gradient $ \nabla_{\matr{A}}\Upsilon=\nabla_{\matr{A}}\Upsilon^{+}-\nabla_{\matr{A}}\Upsilon^{-} $ evaluated with the most recent updates of the factors. Note that this update is always positive if the initial conditions are positive. In practice, to ensure that divisions by zero do not occur the update is modified as follows
\begin{equation}
\hat{\matr{A}}_{k}=\hat{\matr{A}}_{k-1}\boxdot \left[ \nabla_{\matr{A}}\Upsilon_k^{-} \boxslash \text{max}\left(\nabla_{\matr{A}}\Upsilon_k^{+},\varepsilon\boldsymbol{\mathds{1}}\right) \right]
\label{mu_update_2}
\end{equation}
where $ \text{max}(\cdot,\cdot) $ is the element-wise maximum between the elements of the two matrix inputs, $ \varepsilon $ is a small constant and $ \boldsymbol{\mathds{1}} $ is a matrix with ones in all elements. After carrying out both updates of the form (\ref{mu_update_2}) for $ \matr{A} $ and $ \matr{A}' $, we update the other factors in a similar way as in ALS. To solve the scale issue, we also need to normalize $ \hat{\matr{A}}_{k} $, $ \hat{\matr{B}}_{k} $ and $ \hat{\matr{A}}_{k}' $ after each update. Since the factors are positive their columns can be normalized using the $\ell_{1}$ norm, \textit{i.e.} $ \| \vect{a}\|_{1}=\sum_{i} a_{i}  $ instead of the $\ell_{2}$ norm. Also, similarly to the ALS procedure, to start closer to the solution, the coupled MU can be initialized using the results given by two uncoupled MU. In this case, the right permutation may be found by minimizing the $ \beta_{c} $ divergence between the permuted factor $ \hat{\matr{C}}_{0} $ and the fixed factor $ \hat{\matr{C}}'_{0} $.

The MU algorithm is described in Algorithm \ref{mu_alg}, the positive and negative terms in the gradients are given in Appendix \ref{app_1}.

\begin{algorithm}
\caption{MU for positive CP approximation with positive couplings}
\begin{algorithmic}[1]
\REQUIRE {$\tens{Y}$, $\tens{Y}'$, $ \phi $, $ \phi' $, $ \phi_{c} $, $ \beta $, $ \beta' $, $ \beta_{c} $, $ k_{\mathrm{max}} $, $ \Delta\Upsilon_{\text{min}} $, $ \Upsilon_{\text{min}} $}
\STATE Apply two uncoupled MU with random initializations to obtain $ \hat{\matr{A}}_{0} $, $ \hat{\matr{B}}_{0} $, $ \hat{\matr{C}}_{0} $, $ \hat{\matr{A}}_{0}' $, $ \hat{\matr{B}}_{0}' $ and $ \hat{\matr{C}}_{0}' $.
\STATE Permute initialization factors columns to have minimal $ \sum_{ij}d_{\beta_{c}}([\hat{\matr{C}}_{0}]_{ij},\hat{[\matr{C}}_{0}]_{ij}') $. 
\STATE Evaluate the initial cost function $ \Upsilon_{0} $ (\ref{MAP_tweedie}). 
\STATE Set $ k=0 $, $ \Delta\Upsilon=+\infty $.
\WHILE{$ k<k_{\mathrm{max}} $ and $ \Delta\Upsilon_{k}>\Delta\Upsilon_{\text{min}} $,} 
\STATE{k:=k+1.}
\STATE{Update $ \hat{\matr{A}}_{k} $  \\
$ \hat{\matr{A}}_{k}=\hat{\matr{A}}_{k-1}\boxdot \left[ \nabla_{\matr{A}}\Upsilon_k^{-} \boxslash \text{max}\left(\nabla_{\matr{A}}\Upsilon_k^{+},\varepsilon\boldsymbol{\mathds{1}}\right) \right] $}

Normalize columns of $\matr{A}_{k+1}$.
\STATE{Similarly update $\hat{\matr{A}}_{k}'$, $ \hat{\matr{B}}_{k} $, $ \hat{\matr{B}}_{k}' $, $ \hat{\matr{C}}_{k} $ and $ \hat{\matr{C}}_{k}'  $. \\
Normalize $ \hat{\matr{A}}_{k}' $ and $ \hat{\matr{B}}_{k} $, $ \hat{\matr{B}}_{k}' $ remaining unnormalized.}

\STATE{Evaluate the cost function $ \Upsilon_{k} $ (\ref{MAP_tweedie}) and its relative decrease $ \Delta\Upsilon_{k}$.}
\ENDWHILE
\RETURN $ \hat{\matr{A}}_{k} $, $ \hat{\matr{B}}_{k} $, $ \hat{\matr{C}}_{k} $, $ \hat{\matr{A}}_{k}' $, $ \hat{\matr{B}}_{k}' $ and $ \hat{\matr{C}}_{k}' $.
\end{algorithmic}
\label{mu_alg}
\end{algorithm}

\section{Bayesian and hybrid Cram\'er-Rao bounds}\label{Bayesian-sec}
\subsection{Introduction}\label{IntroBayesian-sec}
A lower bound on the MSE can be evaluated using variations of the Cram\'er-Rao bound (CRB). In what follows we will consider two simplifications of the joint Gaussian model, namely the simplified coupling model and unitary first rows for uncoupled factors.

\paragraph*{Simplified coupling model and unitary first rows}in both models we consider Gaussian couplings between components $ \vect{c}_{r} $ and $ \vect{c}_{r}' $ of factors $ \matr{C} $ and $ \matr{C}' $ of the form
\begin{equation}
\vect{c}_{r}=\matr{H} \vect{c}'_{r}  +\boldsymbol{\gamma}_{r},\quad\text{for }r\in\left\lbrace 1,\,,\cdots,\,R \right\rbrace
\label{coupling_CRB}
\end{equation}
where $ \matr{H} $ is a transformation matrix and $ \boldsymbol{\gamma}_{r} $ is a zero mean white Gaussian vector with variance $ \sigma_{c}^{2} $. Similarly to \cite{Liu2001,TichPK13:tsp}, to resolve the scale ambiguity of the CP models, we assume that the components in factors $ \matr{A} $, $ \matr{B} $, $ \matr{A}' $ and $ \matr{B}' $ are of the form $ \vect{a}=\left[1; \;\tilde{\vect{a}}\right]  $, that is, the elements in their first row are known and equal to unity. Without proper constraints on the factors, the information matrices that will be defined later are not invertible and, as a consequence, the bound cannot be evaluated.

\paragraph*{Priors on the factors}we consider a Bayesian model and a hybrid model. They differ in the definition of the priors of all factors except $ \matr{C}' $. In the Bayesian model the unknown part of the components of $ \matr{A} $, $ \matr{B} $, $ \matr{A}' $ and $ \matr{B}' $ have Gaussian priors of the form
\begin{equation}
\tilde{\vect{a}}_{r}=\bar{\vect{a}}_{r}+\boldsymbol{\gamma}_{\vect{a}_{r}}
\label{priors_CRB}
\end{equation}
where $ \bar{\vect{a}}_{r} $ is a constant vector representing the mean of the prior and $ \boldsymbol{\gamma}_{\vect{a}_{r}} $ is a zero mean white Gaussian vector with variance $ \sigma_{\matr{A}}^{2} $. The components of the factor $ \matr{C}' $ have a similar prior, except that the elements in the first row are also unknown.

In the hybrid model, we consider that only $ \matr{C} $ is random, all the other factors are deterministic. 

\paragraph*{Likelihoods} the noise follows the model in (\ref{map_j_gauss}), so that the log-likelihoods can be written as a function of the components in the factors as follows

\small
\begin{equation}
\begin{aligned}
&\mathcal{L}=\log p(\tens{Y};\tilde{\matr{A}},\tilde{\matr{B}},\matr{C})=\\
&-\frac{1}{\sigma_{n}^{2}}\left( \|\vect{y}_{\tilde{\vect{a}}\tilde{\vect{b}}\vect{c}}-\sum\limits_{r=1}^{R} \tilde{\vect{a}}_{r}\kron \tilde{\vect{b}}_{r}\kron \vect{\matr{c}}_{r}\|^{2}+\|\vect{y}_{\tilde{\vect{a}}\vect{c}}-\sum\limits_{r=1}^{R} \tilde{\vect{a}}_{r}\kron \vect{\matr{c}}_{r}\|^{2}\right.\\
&\left.\qquad\qquad\qquad\;+\|\vect{y}_{\tilde{\vect{b}}\vect{c}}-\sum\limits_{r=1}^{R} \tilde{\vect{b}}_{r}\kron \vect{\matr{c}}_{r}\|^{2}+\|\vect{y}_{\vect{c}}-\sum\limits_{r=1}^{R} \vect{\matr{c}}_{r}\|^{2}\right)
\end{aligned}
\label{log_likel_vec}
\end{equation}\normalsize
where $ \vect{y} $ indicates the vectorized version of the data array\footnote{The order of the modes for vectorization is third, second and first mode (order of running indexes). This implies that $ \tenselem{Y}_{ijk}=y_{[(i-1)J+j-1]K+k} $.} \label{vecdef} and the subscripts indicate the free variables underlying the elements in the measurement vector. For example, in $ \vect{y}_{\tilde{\vect{a}}\vect{c}} $ we have all data array elements which do not contain index 1 in the  first mode and contain only index 1 in the second mode.

\begin{table*}[!t]
\begin{tabular}{c|c|c}
 Sub matrices       & $ \matr{F}\times \sigma_{n}^{2} $ & $ \mathbb{E}\left\lbrace  \matr{F} \right\rbrace\times\sigma_{n}^{2} $ \\
 
\hhline{===} $(\tilde{\matr{a}}_{r},\tilde{\matr{a}}_{s})$  & $ (\tilde{\matr{b}}_{r}^{\T}\tilde{\matr{b}}_{s}+1)(\matr{c}_{r}^{\T}\matr{c}_{s})\matr{I} $ & $ [\bar{\matr{b}}_{r}^{\T}\bar{\matr{b}}_{s}+1+(J-1)\sigma_{\matr{B}}^{2}\delta_{rs}]\left\lbrace \bar{\matr{c}}_{r}^{\T}\matr{H}^{\T}\matr{H}\bar{\matr{c}}_{s}+[K\sigma_{c}^{2}+\text{tr}(\matr{H}^{\T}\matr{H})\sigma_{\matr{C}'}^{2}]\right\rbrace \matr{I} $   \\ 

\hline $(\tilde{\matr{a}}_{r},\tilde{\matr{b}}_{s})$  & $ (\matr{c}_{r}^{\T}\matr{c}_{s})(\tilde{\matr{a}}_{s}\kron\tilde{\matr{b}}_{r}^{\T}+\matr{I}) $ & $\left\lbrace \bar{\matr{c}}_{r}^{\T}\matr{H}^{\T}\matr{H}\bar{\matr{c}}_{s}+[K\sigma_{c}^{2}+\text{tr}(\matr{H}^{\T}\matr{H})\sigma_{\matr{C}'}^{2}]\right\rbrace (\bar{\vect{a}}_{s} \kron\bar{\vect{b}}_{r}^{\T}+\matr{I})$   \\ 

\hline $(\tilde{\matr{a}}_{r},\matr{c}_{s})$             & $ (\tilde{\matr{b}}_{r}^{\T}\tilde{\matr{b}}_{s}+1)(\tilde{\matr{a}}_{s}\kron\matr{c}_{r}^{\T}) $ & $[\bar{\matr{b}}_{r}^{\T}\bar{\matr{b}}_{s}+1+(J-1)\sigma_{\matr{B}}^{2}\delta_{rs}](\bar{\matr{a}}_{s}\kron\matr{H}\bar{\matr{c}}_{r}'^{\T})$   \\ 

\hline $(\tilde{\matr{b}}_{r},\tilde{\matr{b}}_{s})$  & $ (\tilde{\matr{a}}_{r}^{\T}\tilde{\matr{a}}_{s}+1)(\matr{c}_{r}^{\T}\matr{c}_{s})\matr{I} $ &$ (\bar{\matr{a}}_{r}^{\T}\bar{\matr{a}}_{s}+1+I\sigma_{\matr{A}}^{2}\delta_{rs})\left\lbrace \bar{\matr{c}}_{r}^{\T}\matr{H}^{\T}\matr{H}\bar{\matr{c}}_{s}+[K\sigma_{c}^{2}+\text{tr}(\matr{H}^{\T}\matr{H})\sigma_{\matr{C}'}^{2}]\right\rbrace \matr{I} $  \\

\hline $(\tilde{\matr{b}}_{r},\matr{c}_{s})$             & $ (\tilde{\matr{a}}_{r}^{\T}\tilde{\matr{a}}_{s}+1)(\tilde{\matr{b}}_{s}\kron\matr{c}_{r}^{\T}) $ &$ [\bar{\matr{a}}_{r}^{\T}\bar{\matr{a}}_{s}+1+(I-1)\sigma_{\matr{A}}^{2}\delta_{rs}](\bar{\matr{b}}_{s}\kron\matr{H}\bar{\matr{c}}_{r}'^{\T})$ \\ 

\hline $(\matr{c}_{r},\matr{c}_{s})$                        & $ (\tilde{\matr{a}}_{r}^{\T}\tilde{\matr{a}}_{s}+1)(\tilde{\matr{b}}_{r}^{\T}\tilde{\matr{b}}_{s}+1)\matr{I} $ & $[\bar{\matr{a}}_{r}^{\T}\bar{\matr{a}}_{s}+1+(I-1)\sigma_{\matr{A}}^{2}\delta_{rs}][\bar{\matr{b}}_{r}^{\T}\bar{\matr{b}}_{s}+1+(J-1)\sigma_{\matr{B}}^{2}\delta_{rs}]\matr{I}$  \\
\hline
\end{tabular} 
\caption{Fisher information and average Fisher information sub matrices for pairs of components.}
\label{tab1}
\end{table*}

\subsection{Bayesian Cram\'er-Rao bound}\label{BayesianCRB-sec}
When all factors are random, we can obtain a bound on the estimation MSE through the Bayesian Cram\'er-Rao bound (BCRB) \cite[pp.4--5]{VanTrees2007}. For a vector of parameters $ \tilde{\theta}=\vec\left(\left[\tilde{\matr{A}};\tilde{\matr{B}};\matr{C};\tilde{\matr{A}}';\tilde{\matr{B}}';\matr{C}'\right]\right) $, we have
\begin{equation}
\text{MSE}(\tilde{\theta}_{i})\geq \mathrm{BCRB}_{ii}
\label{BCR_bound}
\end{equation}
where the $ \mathrm{BCRB} $ is a matrix given by the inverse of a data related information matrix $\matr{F}$ plus a prior related information matrix $\matr{P}$: 
\begin{equation}
\mathrm{BCRB}=\left(\bar{\matr{F}}+\matr{P}\right)^{-1}.
\label{BCRB}
\end{equation}
In what follows, we describe both matrices $ \bar{\matr{F}} $ and $ \matr{P} $. $ \bar{\matr{F}} $ is the average Fisher information matrix given by
\begin{equation}
\bar{\matr{F}}=\left[\begin{array}{cccccc}
 \mathbb{E}[\matr{F}] & \matr{0}     \\ 
 \matr{0}  &  \mathbb{E}[\matr{F}']     
\end{array} \right] 
\label{average_Fisher}
\end{equation}
where the expectation is evaluated w.r.t. the prior distributions. The off-diagonal matrices are equal to zero due to the conditional independence assumption (H1). $ \matr{F} $ is the Fisher information matrix for the first CP model:

\small
\begin{equation}
\matr{F}=\mathbb{E}\left\lbrace \left[ \begin{array}{c}
\nabla_{\tilde{\matr{A}}}\mathcal{L} \\ 
\nabla_{\tilde{\matr{B}}}\mathcal{L} \\ 
\nabla_{\matr{C}}\mathcal{L} 
\end{array} \right] \left[\begin{array}{ccc}
\nabla_{\tilde{\matr{A}}}^{\T}\mathcal{L}& \nabla_{\tilde{\matr{B}}}^{\T}\mathcal{L} & \nabla_{\matr{C}}^{\T}\mathcal{L} \end{array} \right] \right\rbrace 
\label{Fisher}
\end{equation}\normalsize
$ \nabla_{\!\vect{x}}\mathcal{L} $ corresponds to the gradient of the log-likelihoods w.r.t. to a vector $ \vect{x} $. The expectation is evaluated w.r.t. the conditional distribution of the data array given the parameters. The matrix $ \matr{F}' $ for the other CP model is given in a similar way. The gradients w.r.t. components $ \tilde{\vect{a}}_{r} $, $ \tilde{\vect{b}}_{r} $ and $ \vect{c}_{r} $, which are parts of the gradients above, are

\small
\begin{equation}
\begin{aligned}
\nabla_{\tilde{\matr{a}}_{r}}\mathcal{L}=&-\frac{1}{\sigma_{n}^{2} }\left[(\matr{I}\kron\tilde{\vect{b}}_{r}\kron\vect{c}_{r})^{\T}\vect{e}_{\tilde{\vect{a}}\tilde{\vect{b}}\vect{c}}+(\matr{I}\kron\vect{c}_{r})^{\T}\vect{e}_{\tilde{\vect{a}}\vect{c}}\right] \\
\nabla_{\tilde{\matr{b}}_{r}}\mathcal{L}=&-\frac{1}{\sigma_{n}^{2}}\left[ (\tilde{\vect{a}}_{r}\kron\matr{I}\kron\vect{c}_{r})^{\T}\vect{e}_{\tilde{\vect{a}}\tilde{\vect{b}}\vect{c}}+(\matr{I}\kron\vect{c}_{r})^{\T}\vect{e}_{\tilde{\vect{b}}\vect{c}}\right] \\
\nabla_{\matr{c}_{r}}\mathcal{L}=&-\frac{1}{\sigma_{n}^{2}}\left[ (\tilde{\vect{a}}_{r}\kron\tilde{\vect{b}}_{r}\kron\matr{I})^{\T}\vect{e}_{\tilde{\vect{a}}\tilde{\vect{b}}\vect{c}}+(\vect{a}_{r}\kron\matr{I})^{\T}\vect{e}_{\tilde{\vect{a}}\vect{c}}\right.\\
&\qquad\qquad+\left.(\vect{b}_{r}\kron\matr{I})^{\T}\vect{e}_{\tilde{\vect{b}}\vect{c}}+\vect{e}_{\vect{c}}\right] \\
\end{aligned}
\label{grad_log_likel}
\end{equation}\normalsize
where $ \vect{e} $ indicates a vector of the residuals, \textit{e.g.} $ \vect{e}_{\tilde{\vect{a}}\vect{c}}=\vect{y}_{\tilde{\vect{a}}\vect{c}}-\sum_{r=1}^{R} \tilde{\vect{a}}_{r}\kron \vect{\matr{c}}_{r} $, which corresponds here to a Gaussian noise vector. Using the fact that the noise is white and that $ (\vect{x}\kron\vect{y}\kron\vect{z})^{\T}(\vect{x}'\kron\vect{y}'\kron\vect{z}')= (\vect{x}^\T\vect{x}')(\vect{y}^\T\vect{y}')(\vect{z}^\T\vect{z}')$, we can easily evaluate the submatrices in $ \matr{F} $. 
They are given in Table \ref{tab1}, where $ \delta_{rs} $ denotes the discrete delta function. Similar expressions are obtained for $ \matr{F}' $.

Using the coupling and priors models (\ref{coupling_CRB}) and (\ref{priors_CRB}) and the independence hypothesis on the uncoupled factors (H2), we can obtain the average Fisher information matrix $  \mathbb{E}(\matr{F})$, which is also given in Table \ref{tab1}.

The matrix $ \matr{P} $ is the information matrix related to the prior distribution $ p(\tilde{\vect{\theta}}) $, it is given by
\begin{equation}
\matr{P}=\mathbb{E}\left[ \nabla_{\tilde{\vect{\theta}}} \log p(\tilde{\vect{\theta}}) \nabla_{\tilde{\vect{\theta}}}^{\T} \log p(\tilde{\vect{\theta}}) \right]
\label{prior_matrix_1}
\end{equation}
using again the prior models and (H2), we have\small
\begin{equation}
\matr{P}=\left[ \begin{array}{cccccc}
\frac{\matr{I}}{\sigma_{\matr{A}}^{2}} & \matr{0} & \matr{0} & \matr{0} & \matr{0} & \matr{0} \\ 
 \matr{0} & \frac{\matr{I}}{\sigma_{\matr{B}}^{2}} & \matr{0} & \matr{0} & \matr{0} & \matr{0} \\ 
\matr{0} & \matr{0} & \frac{\matr{I}}{\sigma_{\matr{c}}^{2}} & \matr{0} & \matr{0} &  -\frac{\matr{I}\kron\matr{H}}{\sigma_{c}^{2}} \\ 
\matr{0} & \matr{0} & \matr{0} & \frac{\matr{I}}{\sigma_{\matr{A}'}^{2}} & \matr{0} & \matr{0} \\ 
\matr{0} & \matr{0} & \matr{0} & \matr{0} & \frac{\matr{I}}{\sigma_{\matr{B}'}^{2}} & \matr{0} \\ 
\matr{0}  & \matr{0} & -\frac{\matr{I}\kron\matr{H}^{\T}}{\sigma_{c}^{2}} & \matr{0} & \matr{0} & \frac{\matr{I}\kron(\matr{H}^{\T}\matr{H})}{\sigma_{c}^{2}}+\frac{\matr{I}}{\sigma_{\matr{C}'}^{2}}
\end{array}  \right].
\label{prior_matrix_2}
\end{equation}\normalsize
The off-diagonal sub matrices are the effects of coupling between the two CP models on the information matrix to be inverted, the larger is $ \sigma_{c}^{2} $, the smaller is the influence of these sub matrices.

\subsection{Hybrid Cram\'er-Rao bound}\label{hybridCRB-sec}
If we consider the hybrid model where only $ \matr{C} $ is random, a bound on the estimation MSE can be obtained through the hybrid Cram\'er-Rao bound (HCRB) (\cite[p. 12]{VanTrees2007}). The HCRB is given by 
\begin{equation}
\mathrm{HCRB}=\left(\bar{\matr{F}}^{\matr{C}\vert\matr{C}'}+\matr{P}^{\matr{C}\vert\matr{C}'}\right)^{-1}.
\label{HCRB}
\end{equation}
Matrices $ \bar{\matr{F}}^{\matr{C}\vert\matr{C}'} $ and $ \matr{P}^{\matr{C}\vert\matr{C}'} $ play similar roles as $ \bar{\matr{F}} $ and $ \matr{P} $ for the BCRB. The matrix $ \bar{\matr{F}}^{\matr{C}\vert\matr{C}'} $ is evaluated in a similar way as $ \bar{\matr{F}} $ with the difference that the expectation is taken w.r.t. the conditional distribution of the coupled factor $ p(\matr{C}\vert\matr{C}') $ only, that is 
\begin{equation}
\bar{\matr{F}}^{\matr{C}\vert\matr{C}'}=\left[\begin{array}{cccccc}
 \mathbb{E}_{\matr{C}\vert\matr{C}'}[\matr{F}] & \matr{0}     \\ 
 \matr{0}  &  \matr{F}'     
\end{array} \right].
\label{hybrid_Fisher}
\end{equation}
The submatrix $ \matr{F}' $ is directly the Fisher information for the second CP model and the elements in $ \mathbb{E}_{\matr{C}\vert\matr{C}'}[\matr{F}] $ are similar to the elements in $ \matr{F} $, only the components $ \matr{c}_{r} $ are affected by the expectation. For the hybrid model $ \mathbb{E}[\vect{c}_{r}]=\vect{c}_{r}' $ and $ \mathbb{E}[\vect{c}_{r}^{\T}\vect{c}_{s}]=\vect{c}_{r}'^{\T}\matr{H}^{\T}\matr{H}\vect{c}_{s}+K\sigma_{c}^{2}\delta_{rs} $.

The prior matrix is evaluated w.r.t. the distribution of the random parameters $ \tilde{\vect{\theta}}^{r} $ conditioned on the non random parameters $ \tilde{\vect{\theta}}^{nr} $
\begin{equation}
\matr{P}^{\matr{C}\vert\matr{C}'}=\mathbb{E}_{\tilde{\vect{\theta}}^{r}\vert \tilde{\vect{\theta}}^{nr}}\left[ \nabla_{\tilde{\vect{\theta}}} \log p(\tilde{\vect{\theta}}^{r}\vert \tilde{\vect{\theta}}^{nr}) \nabla_{\tilde{\vect{\theta}}}^{\T} \log p(\tilde{\vect{\theta}}^{r}\vert \tilde{\vect{\theta}}^{nr}) \right].
\label{prior_hybrid_1}
\end{equation}
For the hybrid coupled CP model we have\small
\begin{equation}
\matr{P}=\left[ \begin{array}{cccccc}
\matr{0} & \matr{0} & \matr{0} & \matr{0} & \matr{0} & \matr{0} \\ 
 \matr{0} & \matr{0} & \matr{0} & \matr{0} & \matr{0} & \matr{0} \\ 
\matr{0} & \matr{0} & \frac{\matr{I}}{\sigma_{\matr{c}}^{2}} & \matr{0} & \matr{0} &  -\frac{\matr{I}\kron\matr{H}}{\sigma_{c}^{2}} \\ 
\matr{0} & \matr{0} & \matr{0} & \matr{0} & \matr{0} & \matr{0} \\ 
\matr{0} & \matr{0} & \matr{0} & \matr{0} & \matr{0} & \matr{0} \\ 
\matr{0}  & \matr{0} & -\frac{\matr{I}\kron\matr{H}^{\T}}{\sigma_{c}^{2}} & \matr{0} & \matr{0} & \frac{\matr{I}\kron(\matr{H}^{\T}\matr{H})}{\sigma_{c}^{2}}
\end{array}  \right].
\label{prior_matrix_2_hybrid}
\end{equation}\normalsize
We can still see the off diagonal sub matrices indicating the coupling, but parts of the diagonal elements are now equal to zero, which means that the hybrid model is less regularized than the Bayesian model.


\section{Simulations}\label{simulations-sec}
In this section we simulate the algorithms for estimating the latent factors of the coupled CP models in different settings\footnote{MATLAB\textsuperscript{\textregistered} codes related to the simulations reported in this section can be found at: 
\href{http://www.gipsa-lab.fr/~pierre.comon/TensorPackage/tensorPackage.html}{www.gipsa-lab.fr/~pierre.comon/TensorPackage}. }. 
We focus mainly on the hybrid case where one coupled factor is random and all other factors are unknown but considered to be deterministic. First we apply the ALS algorithm (Algorithm \ref{als_alg}) to two cases of the joint Gaussian model: direct coupling of the $ \matr{C} $ factors and coupling of one component. In the direct coupling case, we compare the results with the HCRB. Then we show some simulation results on the compressed coupled setting. We simulate also the performance of Algorithm \ref{mu_alg} on the estimation of a Gamma-coupled positive Gamma model.  At the end of the simulation section, we return to the joint Gaussian model and simulate the ALS algorithm in the case when  the $ \matr{C} $ factors are coupled through interpolation. 
Note that the coupling standard deviation $\sigma_c$  in (\ref{hybridgaussian}) is not considered to be a nuisance parameter in these simulations since the norms of the columns of the coupled factors are known through normalization of the columns of other factors to one.
 
All simulation results are given in terms of a Monte-Carlo estimation of the total MSE of one or a few factors. The total MSE on a factor, for example $ \matr{C} $, based on $ N_{r} $ different noise realizations is $ (1/N_{r})\sum_{k=1}^{K}\sum_{r=1}^{R}\sum_{n=1}^{N_{r}}(C_{kr}-\hat{C}_{kr}^{n})^{2} $, where $ \hat{C}_{kr}^{n} $ is the factor estimated in the $ n-th $ noise realization. The expressions used to evaluate the different signal to noise ratios (SNR) in the simulations are given in the Appendix \ref{app_2}.

\subsection{Similar factors}\label{similarFactors-sec}
\begin{figure}
  \centering
  \centerline{\includegraphics[width=7cm]{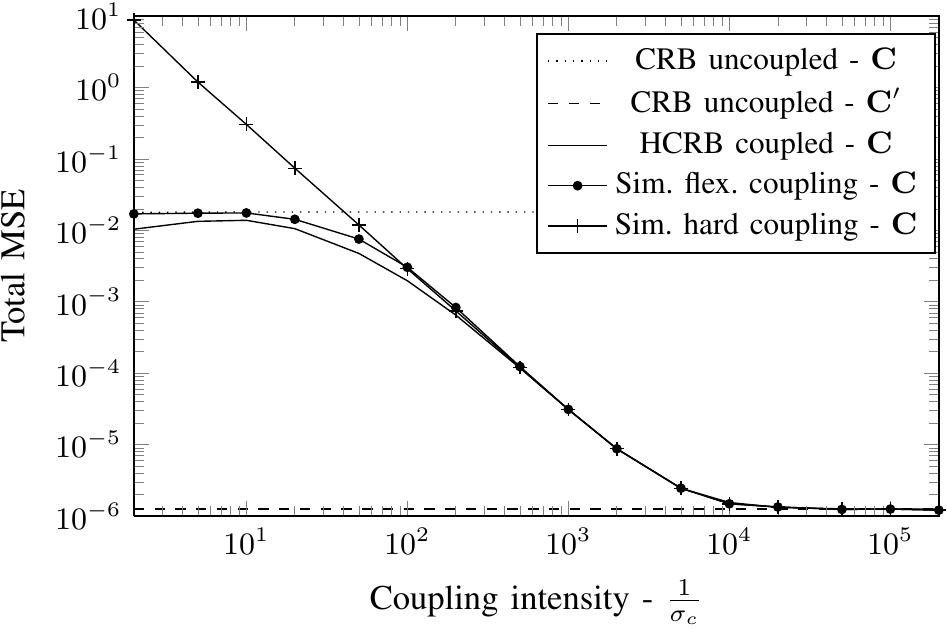}}
\caption{Total MSE for the factors $ \matr{C} $ and $ \matr{C}' $ of the coupled CP model as a function of the coupling intensity $ \frac{1}{\sigma_{c}} $. Results are shown for ALS algorithms considering flexible couplings (Algorithm \ref{als_alg}) and hard couplings ($ \matr{C}=\matr{C'} $). The CP models are measured with different noise levels $ \sigma_{n}=0.1 $ and $ \sigma_{n}'=0.001 $.}
\label{fig1}
\end{figure}

We start with a straightforward hybrid Gaussian coupling model $ \matr{C}=\matr{C}'+\matr{\Gamma} $. The two CP models are generated randomly (i.i.d. Gaussian $ \mathcal{N}\left(0,1\right) $) with dimensions $ I=I'=J=J'=K=K'=10 $ and $ R=R'=3 $. The first rows in the factors $ \matr{A} $, $ \matr{B} $, $ \matr{A}' $ and $\matr{B}$ are set to $\vect{1}$, so that we can compare the results with the corresponding HCRB. Here the columns of factor matrices are thus not normalized. We vary the coupling intensity $\frac{1}{\sigma_{c}} $ from $ 2 $ to $ 2\times 10^{5} $. 
The data array $ \tens{Y}$ is almost noiseless $ \text{SNR}(\tens{Y})\in\left[64.77;65.74\right] \text{ dB} $ ($ \sigma_{n}=0.001 $ (\ref{SNR1})), while $ \tens{Y}' $ has some noise $ \text{SNR}(\tens{Y}')= 24.77 \text{ dB} $ ($ \sigma_{n}'=0.1$ (\ref{SNR2})). Algorithm \ref{als_alg} is applied to estimate the CP models under $ 500 $ different noise and coupling realizations. The total MSE on the $ \matr{C} $ and $ \matr{C}' $ factors is evaluated. We also evaluate the total MSE for an ALS algorithm using hard couplings, \textit{i.e.} $ \matr{C}=\matr{C}' $ and the uncoupled CRB for both CP models. Since $ \tens{Y} $ is generated with different $ \matr{C} $ due to the random coupling, we average the resulting CRB.  The results are shown in Fig.\ref{fig1}.

We can see that when the coupling is weak the MSE on $ \matr{C} $ is close to its uncoupled CRB. An intuitive explanation for the behaviour presented in the figure is that, by increasing the coupling intensity, the factor is estimated with a much better performance, since more information comes from the clean tensor through the coupling. The flexible coupling allows to assess the continuous transition between uncoupled models and exactly coupled models. The HCRB predicts well this transition when the factors are strongly coupled, however a gap is present when the models are uncoupled. Moreover in the interval $ 1/\sigma_{c}\in[10;100] $, a flexible coupling works better than a hard coupling. For $ 1/\sigma_{c}<10 $, the performance is similar to the uncoupled case, while applying a hard coupling in this case will lead to very poor results. For $ 1/\sigma_{c}>100 $, flexible couplings and hard couplings present equivalent performance. Note that in Fig. \ref{fig1}, the HCRB for the clean factor  practically coincides with the ``CRB uncoupled $\matr{C}'$'' and is hence not displayed. 

\subsection{Shared component}\label{sharedComponent-sec} 
\begin{figure}
  \centering
  \centerline{\includegraphics[width=7cm]{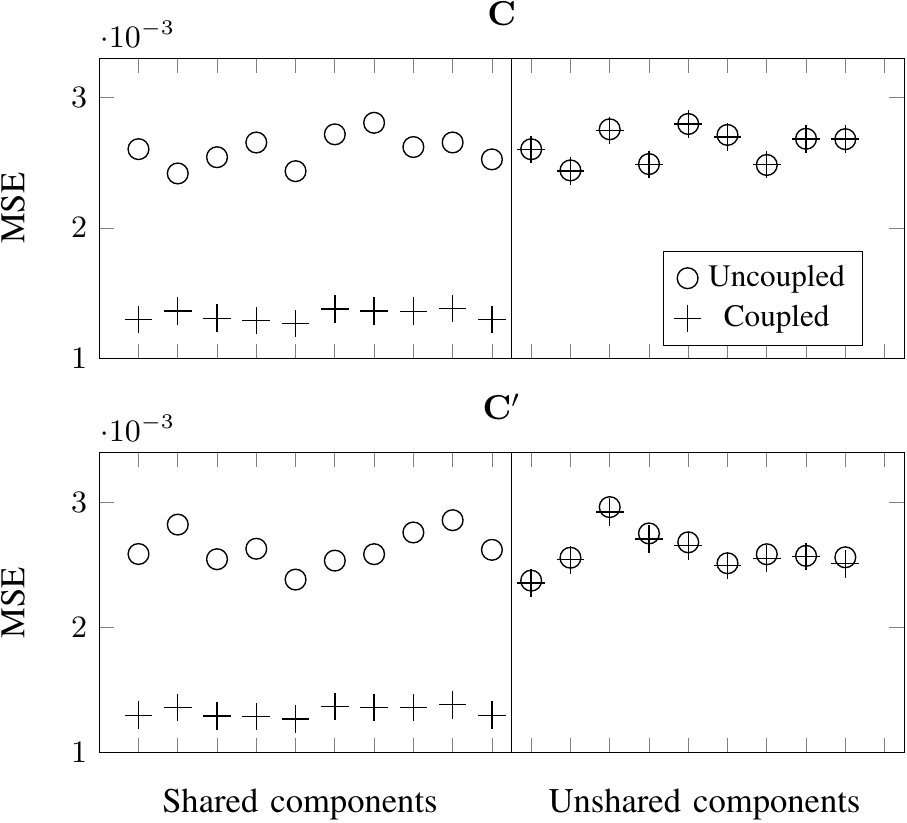}}
\caption{MSE for the elements of $ \matr{C} $ and $ \matr{C}' $ of the coupled CP models. The models have one similar component (a shared component) on their $ \matr{C} $ factors while the other component is not similar.}
\label{fig2}
\end{figure}

We also simulate the case when components $ \vect{c}_{1} $ and $ \vect{c}_{1}' $ are shared between the models and the noise levels are similar. In this case $ I=I'=J=J'=K=K'=10 $, $ R=R'=2 $ and $ \sigma_{c}=0.001 $. Factors $ \matr{A} $, $ \matr{B} $, $ \matr{A}' $, $ \matr{B}' $ and $\matr{C}'$ are drawn coefficient-wise from standard Gaussian distributions, then normalized column-wise with the $\ell_2$ norm except for $\matr{C}$ and $\matr{C}'$, and $ \text{SNR}(\tens{Y})\approx\text{SNR}(\tens{Y}')=9 \text{ dB} $ ($ \sigma_{n}=0.05 $ and $ \sigma_{n}'=0.05 $ (\ref{SNR3})). The MSE on all elements of both $ \matr{C} $ and $ \matr{C}' $ are evaluated in a similar way as above. The MSE for the estimation of each element of the factors is shown in Fig. \ref{fig2}.

Although the unshared component is not improved by the coupling, we observe in Fig. \ref{fig2} that the MSE is halved when double the amount of data is used (for the shared components). We might expect this behaviour when we are solving approximately a linear system.

\subsection{Compressed coupled CP}\label{compressedCoupledCP-sec}

\begin{figure}[t]
  \centering
  \centerline{\includegraphics[width=7cm]{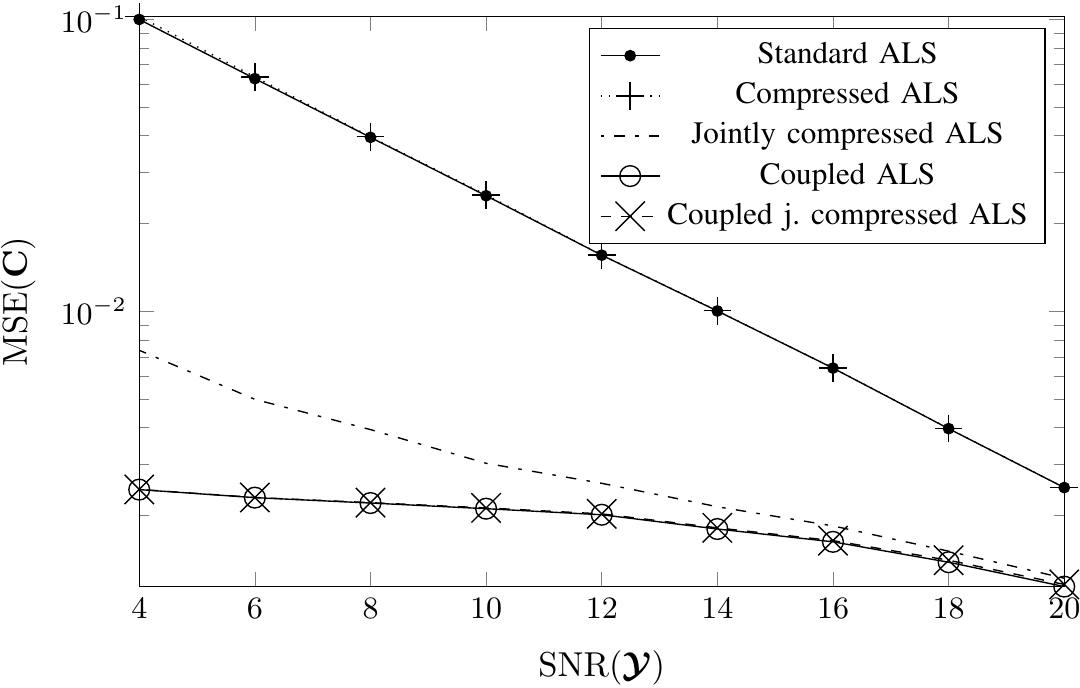}}
\caption{Reconstruction MSE of noisy factor $\matr{C}$ with compressed coupled algorithms. Comparison is made with uncoupled and uncompressed algorithms.}
\label{fig3}
\end{figure}

For large coupled data sets, we simulate the joint compression strategy on $100\times100\times100$ tensors\footnote{Larger sizes could have been used but they would require very long Monte-Carlo simulations, since we compare the performance with the uncompressed strategy.}, $R=5$ and set multilinear rank guesses to $R_1=R_2=R_3=5$ for both coupled tensors. Factors are drawn randomly similarly to the previous example. $\matr{C}$ is defined from unnormalized $\matr{C}'$ with additive zero mean Gaussian noise of variance $\sigma_c=10^{-3}$. The SNR for $\tens{Y}'$ is set to $22\text{ dB}$ ($ \sigma_{n}'=0.0018 $ (\ref{SNR4})), and it varies from $4 \text{dB}$ to $20 \text{ dB}$ for $\tens{Y}$ ($ \sigma_{n}\in[0.0141;0.0022] $ (\ref{SNR3})). We compare the performance of the coupled estimation in the compressed space with the coupled approach in the uncompressed space and with three uncoupled approaches: standard ALS in the uncompressed space and ALS in the compressed space with independent and joint compression. Results for $N=50$ noise realizations and $50$ iterations of the coupled algorithms are shown in Fig. \ref{fig3}. Compression was computed with three randomized SVD from \cite{Halko2011}. Moreover, initializations were given by two uncoupled uncompressed ALS with 1000 iterations, themselves randomly initialized.

As expected, compression decreases estimation performance in both coupled and uncoupled cases, even more as the noise level increases. But it is worth noting that even coupling only in the compression improves performance w.r.t. simply running an independent ALS on $\tens{Y}$. Also, the joint-compressed coupled case behaves well. On the other hand, compressing dramatically improves the runtime of joint decompositions which can be quite slow. Table \ref{tab2} shows a clear difference in computation time\footnote{Computational time is shown for a MATLAB\textsuperscript{\textregistered} implementation of the algorithms. The algorithms are tested on a 3.2 GHz Intel\textsuperscript{\textregistered} Core\textsuperscript{TM} i5 with 8Gb of RAM.}. Also, compressing allows to go to much higher dimensions than the ones used in these simulations. 

\begin{table}[h]\small 
\begin{tabular}{lcc}
 & Joint-Compressed & No Compression \\
\hhline{===}
Time $(s)$ &  $0.04$ & $0.92$ \\

Standard deviation $(s)$ & $ <0.01 $ & $0.08$ 
\end{tabular}\normalsize
\caption{Mean runtime for 50 iterations of the joint-compressed coupled decomposition algorithm and the coupled algorithm without compression. Mean is taken over all realizations and all noise levels.}
\label{tab2}  
\end{table}

\subsection{Gamma coupling of positive variables}\label{gammaCoupling-sec}
In the case of positive variables, we simulate the coupling of positive CP models with all dimensions equal to $ 10 $ and $ R=3 $. Both the measurement and coupling models follow a Gamma distribution, that is $ \beta=\beta'=\beta_{c}=2 $ in the Tweedie's distribution, with dispersion parameters $ \phi=0.5 $, $ \phi'=0.05 $ and $ \phi_{c}=0.05 $\footnote{Note that here defining a SNR is difficult since the noise power depends on the signal power.}. We generate the CP models randomly. The elements in the factors are the absolute values of i.i.d. Gaussian variables $ \mathcal{N}(0,1) $ and the columns of $ \matr{A} $, $ \matr{B} $, $ \matr{A}' $ and $ \matr{B}' $ are normalized with their $ \ell_{1} $ norms. Both uncoupled and coupled (Algorithm \ref{mu_alg}) algorithms are applied to $ 50 $ different realizations of the noise distribution, in this case multiplicative noise. We impose the constraint that $ 2 $ components in $ \matr{A} $ are nearly collinear (correlation coefficient of $ 0.99997 $). The total MSE for the factors are shown in Table \ref{tab3}.
\begin{table}[h]\small
\begin{tabular}{c|cccccc}
 & $\matr{A}$ & $\matr{B}$ & $\matr{C}$ & $\matr{A}'$ & $\matr{B}'$ & $\matr{C}'$\\
\hhline{=======}
Uncoupled & $0.041$ & $0.054$ & $4.904$ & $0.001$ & $0.001$ & $0.129$ \\
Coupled & $0.015$ & $0.021$ & $0.803$ & $0.001$ & $0.001$ & $ 0.127 $
\end{tabular}\normalsize
\caption{Total MSE for the estimation of the factors in the coupled CP models with Gamma likelihoods and couplings. Two components in $ \matr{A} $ are nearly collinear. The results are shown for both uncoupled and coupled (Algorithm \ref{mu_alg}) MU algorithms. }
\label{tab3}
\end{table}

The results show a large gain in performance in all factors in the noisy tensor. The fact that $ \matr{A} $ has nearly collinear columns influences strongly the conditioning of the uncoupled problem. In the Gamma coupled case, information flows from the clean tensor to the noisy tensor in an heterogeneous way. The elements in $ \matr{C}' $ are linked to the variations in $ \matr{C}' $, since the model is multiplicative. This heterogeneous information flow allows to better estimate $ \matr{C} $ and, since the model is ill conditioned, a reduction on the estimation error in $ \matr{C} $ will greatly affect the estimation performance of $ \matr{A} $ and $ \matr{B} $.

\subsection{Different sampling rates}\label{differentSampling-sec}
As another non trivial example of coupling we consider the case when $ I=I'=J=J'=10 $, but with different sizes on the third mode $ K=37 $ and $ K'=53 $. We suppose that the components on $ \matr{C} $ and $ \matr{C}' $ are uniformly sampled versions of the same underlying continuous functions, however, the sampling periods to obtain the factors are different so that the elements in the factors cannot be similar, at least, in most of the points. Since the factors are not similar, we cannot apply the direct coupling model and we must use an interpolation approach as explained at the end of Sec. \ref{sec:examples}.

To obtain the interpolation kernel we assume that the sampled functions are band-limited and periodic. For an odd number of samples, the interpolation kernel is given by the Dirichlet kernel \cite{Margolis2008}\small
\begin{equation}
 \matr{H}_{lk}=\frac{\sin\lbrace K\pi[t_l-t_k]/[(L-1) T^{j}\rbrace}{K\sin\lbrace\pi[t_l -t_k]/[(L-1) T^{j}\rbrace}
\label{samp_kernel}
\end{equation}\normalsize
where $ t_l=(l-1) T^{j} $ with $ l={1,\,\ldots,\,L} $ is the uniform interpolation grid with interpolation period $ T^{j} $, $ t_k=(k-1) T $ with $ k={1,\,\ldots,\,K} $ and $ t_k'=(k'-1) T '$ with $ k'={1,\,\ldots,\,K'} $ are the original sampling grids with sampling periods $ T $ and $ T' $.

\begin{figure}
  \centering
  \centerline{\includegraphics[width=7cm]{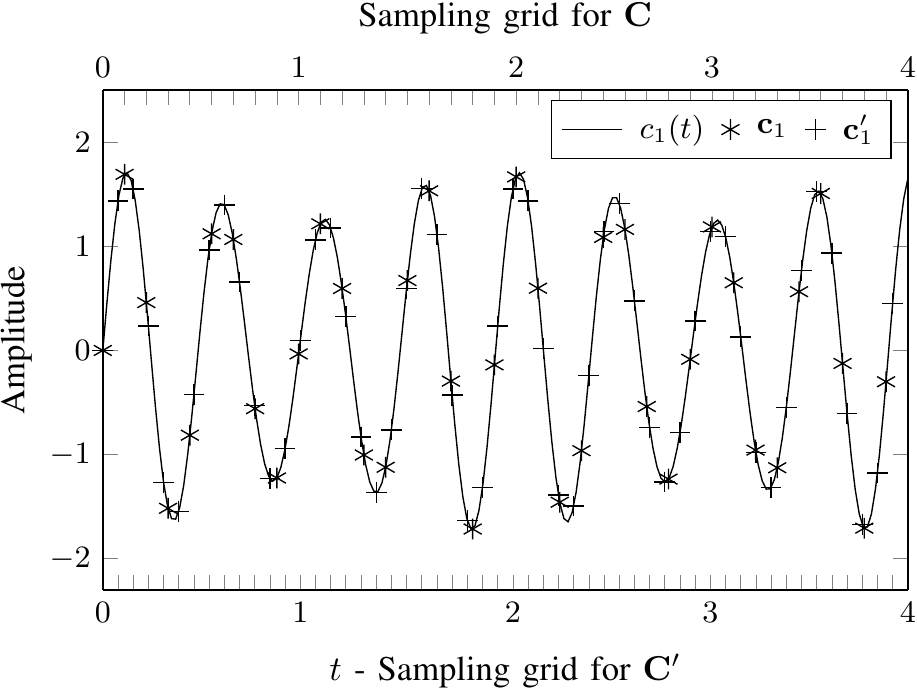}}
\caption{Underlying continuous function $ c_{1}(t) $ for the first components of the $ \matr{C} $ factors and their corresponding sampled versions $ \vect{c}_{1} $ and $ \vect{c}_{1}' $ obtained with different sampling grids.}
\label{fig4}
\end{figure}

We simulate two random CP models, $ R=R'=3 $ and normalize factors $ \matr{A} $, $ \matr{B} $, $ \matr{A}' $ and $ \matr{B}' $. The components on the $ \matr{C} $ factors are generated by regularly sampling $ c_{r}(t)=\sum_{i=1}^{N_{c}} \alpha_{ir}\sin(2\pi f_{i} t) $ in the time interval $ t\in[0;4] $ where $ \alpha_{ir} $ are drawn from i.i.d. Gaussian variables $ \mathcal{N}(0,1) $. We assume $ N_{c}=3 $ and $ f_{1}=2.05 $, $ f_{2}=2.55 $, $ f_{3}=3.5 $. Observe that the first two components do not have an integer number of periods in the considered time duration, therefore these components can be only approximated and not exactly reconstructed with the assumed kernel. The sampling periods $ T=4/37 $, $ T'=4/53 $ are chosen to have $ K $ and $ K' $ samples in the given time duration. An example of continuous-time component with its sampled points on different grids is shown in Fig. \ref{fig4}.

\begin{figure}[t!]
\centering
\centerline{\includegraphics[width=7cm]{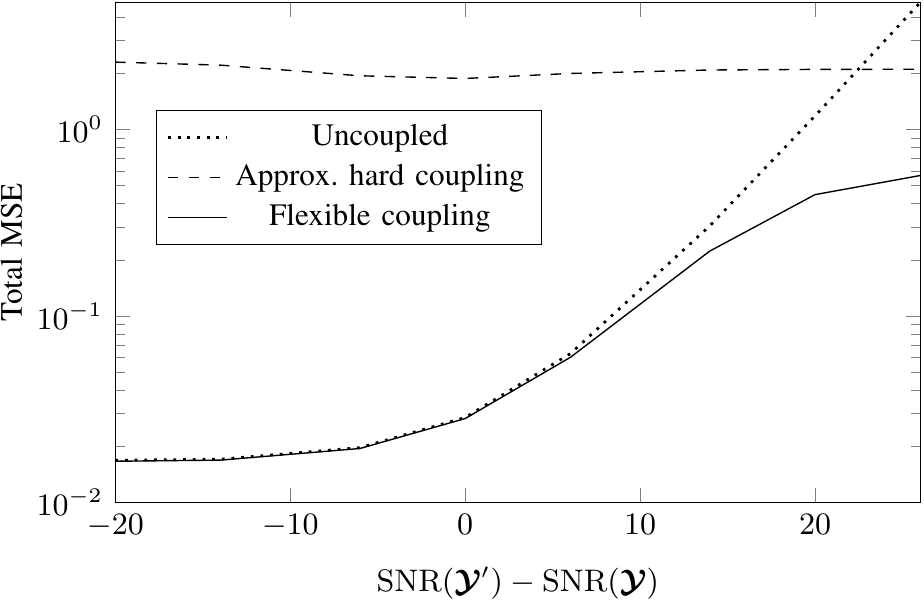}}
\caption{Total MSE for the estimation of the $ \matr{C} $ factors for different $ \text{SNR}(\tens{Y}')-\text{SNR}(\tens{Y}) $. The CP models are coupled through interpolation. Estimation is carried out considering three different models: uncoupled, approximately hard coupled ($ \sigma_c=10^{-4} $) and flexibly coupled ($ \sigma_c=0.15 $).}
\label{fig5}
\end{figure}

We fix $ \text{SNR}(\tens{Y}')=26.46\text{ dB}$ ($\sigma_{n}'=0.01 $ (\ref{SNR5})), while $ \text{SNR}(\tens{Y})$ varies from $ 6.46\text{ dB} $ to $ 46.46\text{ dB} $ ($ \sigma_{n}\in[0.001;0.1] $ (\ref{SNR5})) so that $ \text{SNR}(\tens{Y}')-\text{SNR}(\tens{Y})\in\left[-20\,;\,20\right]  $. For the considered model, interpolation can only approximate the continuous-time signal. Therefore, it is necessary to set a nonzero $ \sigma_{c} $ even if the continuous signals are the same for both data sets. We set $ L=100 $ ($ T^{j}=4/100 $), $ \sigma_{c}=0.15 $ and we evaluate the total MSE on the coupled factors with $ 200 $ noise realizations in the same way as before. We also evaluate the total MSE for approximately hard couplings ($ \sigma_c=10^{-4} $). The results are shown in Fig. \ref{fig5}. When the noise ratio increases, the total MSE for the uncoupled approach increases sharply, while the coupled approach has a smooth increase. This shows that even though the coupled factors are not similar, information can still be exchanged between them through interpolation. We can also see that an approximately hard coupling approach does not lead to good results in this case.

\begin{figure}[t!]
\centering
\centerline{\includegraphics[width=7cm]{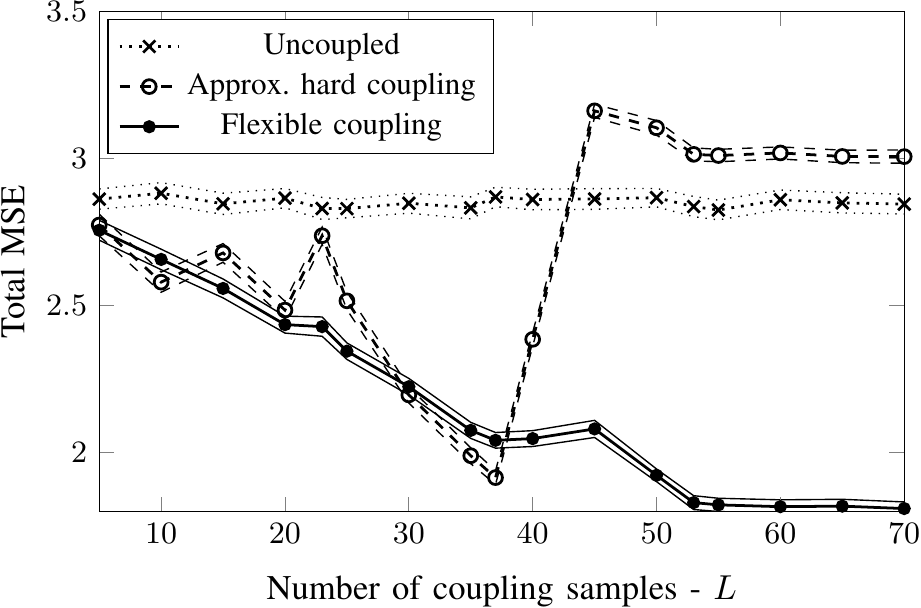}}
\caption{Total MSE for the estimation of the $ \matr{C} $ factors as a function of the number of interpolation samples. Estimation is carried out considering three different models: uncoupled, approximately hard coupled model ($ \sigma_c=10^{-4} $) and flexibly coupled ($ \sigma_c=0.15 $). The lines above and below each plot indicate $ 95\% $ confidence intervals for the estimation of the total MSE.}
\label{fig6}
\end{figure}

We consider also the case when both data arrays are noisy $ \text{SNR}(\tens{Y'}) \approx \text{SNR}(\tens{Y})=6.46 \text{ dB}  $ ($ \sigma_{n}=\sigma_{n}'=0.1$ (\ref{SNR5})) and the number of interpolation points for the coupling $ L $ varies from $ 5 $ to $ 75 $. We use the previously considered coupling parameters $ \sigma_c $ for the flexible and approximately hard coupling approaches. Total MSE values obtained with $ 200 $ different realizations of noise are shown in Fig. \ref{fig6}.

By increasing the number of interpolation points the information exchanged within the model is larger and the MSE decreases almost linearly for the flexible coupling approach. Above $ L=53 $, only a small quantity of information can be exchanged because this is the maximum resolution present in the data and the total MSE curve becomes almost flat. The approximately hard coupling approach gives its best result at coarsest resolution $ L=37 $, which is better than flexible coupling for this particular value of $ L $ but worse than the best results for the flexible coupling approach ($ L>53 $). Observe, as indicated by the confidence intervals in the figure, that the total MSE oscillations seem to be linked to a deterministic pattern and not to the specific set of noise realizations.

As a last simulation we consider the case when one tensor is clean $ \text{SNR}(\tens{Y})=46.67\text{ dB} $ ($ \sigma_n=0.001 $ (\ref{SNR5})) and is of size $ 10\times10\times 24 $ with a sampling period $ T=4/24 $ in the third mode. The other tensor is very noisy $ \text{SNR}(\tens{Y})=-5.61\text{ dB} $ ($ \sigma_n=0.001 $ (\ref{SNR5})), but it has a higher resolution in the third mode, its size is $ 10\times10\times 37 $ and sampling period $ T'=4/37 $. The frequencies in the components are $ f_{1}=3.22 $, $ f_{2}=3.47 $, $ f_{3}=3.73 $, as a consequence, the first tensor cannot be used to recover the continuous components in the third mode since its sampling frequency is below $ 2f_{1} $.

We cannot couple the CP models using interpolation on both factors, since the continuous-time components cannot be reconstructed from the factor $ \matr{C} $, however we can interpolate the high resolution $ \matr{C}' $ to have the same low resolution grid as $ \matr{C} $ and then couple them using a flexible coupling ($ \sigma_c=0.15 $). We simulated this approach using uncoupled and coupled algorithms. The algorithms are applied to $ 200 $ noise realizations. After retrieving factors $ \matr{C} $ and $ \matr{C}' $, we interpolate both factors $ \matr{C} $ and $ \matr{C}' $ using the sampling kernel (\ref{samp_kernel}) with a very fine grid (5000 points) to simulate the continuous function and we integrate numerically the squared error (using trapezoidal integration), so that we have a total squared error which approximates the integral of the continuous squared error. We then average the total squared errors. The result is given in Table \ref{tab4}.
\begin{table}[h]
\begin{tabular}{ccc}
 Algorithm & Factor & Total squared error \\ \hhline{===}
 Uncoupled & $ \matr{C} $ & $33.491$ \\ 
 Coupled & $ \matr{C} $ &  $33.494$ \\ 
 Uncoupled & $ \matr{C}' $ &  $3.959$ \\ 
 Coupled & $ \matr{C}' $ &   $1.061$
\end{tabular} 
\label{tab4}
\caption{Total MSE for the reconstruction of the continuous band-limited components from factors $ \matr{C} $ and $ \matr{C}' $ using uncoupled and coupled models.}
\end{table}

Although the factor $ \matr{C} $ does not have enough information to recover within some acceptable accuracy the underlying continuous function, with or without the coupling, it still contains a lot of information which can be transferred through the interpolation coupling. The information contained in $ \matr{C} $ allows to better estimate $ \matr{C}' $ and, as a consequence, to better reconstruct the continuous function. 

\section{Conclusions}
Since the expression of a phenomenon can be different in various data sets, the link between factorizations of the data sets should be somehow flexible. To give a meaning to this flexibility we have proposed in this paper a Bayesian setting for the coupling between factors of the tensor CP model, which encompasses matrix-matrix and matrix-tensor couplings. Under this setting, we can explore multimodal data fusion problems by proposing not only trivial flexible links between factors, \textit{e.g.} i.i.d. Gaussian models for the differences between factors having common dimensions, but also joint Gaussian models in transformed domains, sparse similarity models and positive similarity models. 

We proposed algorithms to tackle the underlying optimization problems, taking into account scaling ambiguities. The latter turn out to play a key role in coupled models, and needed a special care. To allow scalability to large datasets, a compression scheme for coupled data has been derived; this scheme is being extended to deal with more general coupling models. Finally, we provided extensive applications of the Bayesian framework and evaluated estimation performances w.r.t. the computed CRB for both hybrid and joint Gaussian models. We found that coupling can help improving the conditioning of ill-conditioned problems, and improves estimation accuracy overall. In a large number of applications, the flexibility offered by our coupling approach has revealed to be of crucial importance.

In this work we have focused on the MAP estimation of the latent factors, where all parameters of the coupling model are considered to be known. Also, to retrieve the factors, we proposed modifications of well-known easy to implement algorithms. Future directions of work in this subject may include MMSE estimation of parameters using Markov chain Monte Carlo techniques, full Bayesian approaches to estimate the parameters of the coupling model when they are unknown, evaluation of the effect of the flexible coupling models on tensor completion and comparison of performance of different optimization approaches.


\appendices

\section{Warm start and cold start}
\label{app_3}
\begin{figure}[t!]
\centering
\centerline{\includegraphics[width=7cm]{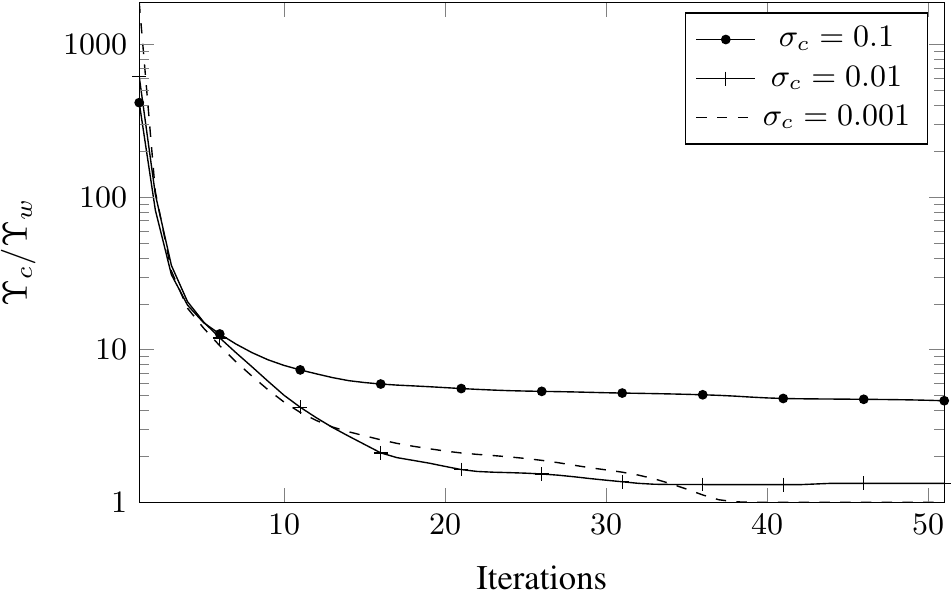}}
\caption{Mean of the ratio between the cost function values for the coupled CP ALS using random initialization (cold start) $ \Upsilon_{c} $ and the results given by uncoupled ALS (warm start) $ \Upsilon_{w} $.}
\label{fig7}
\end{figure}

Here we give some simulations to illustrate the gap in performances between coupled algorithms initialized with two independent ALS (or any other classical tensor algorithm) and two random guesses. The experiments were done for a Gaussian coupling model of the type $ \matr{C}=\matr{C}'+\matr{\Gamma} $. The ratio of the averaged cost functions are shown in Fig. \ref{fig7} for different $ \sigma_{c} $. The tensor dimensions are $ I=J=K=I'=J'=K'=10 $ and $ R=3 $. The noise levels are $ \sigma_{n}=0.03 $ and $ \sigma_{n}=0.01 $ and the number of realizations is $ 100 $. Using warm start seems better than cold start. In practice, a much larger number of swamps has been observed in cold start than in warm start.

%
%

\section{Gradient matrices for the MU algorithm}
\label{app_1}
The negative and positive terms of the gradient matrix of cost function (\ref{MAP_tweedie}) w.r.t. factor $ \matr{A} $ are\small
\begin{equation}
\begin{aligned}
\nabla\Upsilon_{\matr{A}}^{-}=&\frac{1}{\phi}\left\lbrace \matr{Y}_{(1)}\boxdot\left[\matr{X}_{(1)}\right]^{-\beta}\right\rbrace \left(\matr{C}\khatri\matr{B} \right) \\
\nabla\Upsilon_{\matr{A}}^{+}=&\frac{1}{\phi}\left[\matr{X}_{(1)}\right]^{1-\beta}\left(\matr{C}\khatri\matr{B} \right)
\end{aligned}
\label{eq_grad_A}
\end{equation}\normalsize
where $ \matr{X}_{(1)}=\matr{A}\left(\matr{C}\khatri\matr{B} \right)^{\T} $ is the first unfolding of the CP model $ \tens{X} $. For the factor $ \matr{B} $ we have\small
\begin{equation}
\begin{aligned}
\nabla\Upsilon_{\matr{B}}^{-}=&\frac{1}{\phi}\left\lbrace \matr{Y}_{(2)}\boxdot\left[\matr{X}_{(2)}\right]^{-\beta}\right\rbrace \left(\matr{C}\khatri\matr{A} \right) \\
\nabla\Upsilon_{\matr{B}}^{+}=&\frac{1}{\phi}\left[\matr{X}_{(2)}\right]^{1-\beta}\left(\matr{C}\khatri\matr{A} \right)
\end{aligned}
\label{eq_grad_B}
\end{equation}\normalsize
where $ \matr{X}_{(2)}=\matr{B}\left(\matr{C}\khatri\matr{A} \right)^{\T} $. For factor $ \matr{C} $ an additional term appear due to the coupling\small
\begin{equation}
\begin{aligned}
\nabla\Upsilon_{\matr{C}}^{-}=&\frac{1}{\phi}\left\lbrace \matr{Y}_{(3)}\boxdot\left[\matr{X}_{(3)}\right]^{-\beta}\right\rbrace \left(\matr{B}\khatri\matr{A} \right)+\frac{\left[\matr{C}\right]^{1-\beta_{c}}}{\phi_{c}(\beta_{c}-1)} \\
\nabla\Upsilon_{\matr{C}}^{+}=&\frac{1}{\phi}\left[\matr{X}_{(3)}\right]^{1-\beta}\left(\matr{B}\khatri\matr{A} \right)+  \frac{\beta_{c}}{2}\left(\mathds{1}\boxslash\matr{C}\right)+\frac{\left[\matr{C}'\right]^{1-\beta_{c}}}{\phi_{c}(\beta_{c}-1)}.
\end{aligned}
\label{eq_grad_C}
\end{equation}\normalsize
The gradients w.r.t. $ \matr{A}' $ and $ \matr{B}' $ have the same form as (\ref{eq_grad_A}) and (\ref{eq_grad_B}), while the gradient w.r.t. $ \matr{C}' $ is given by\small
\begin{equation}
\begin{aligned}
\nabla\Upsilon_{\matr{C}'}^{-}=&\frac{1}{\phi'}\left\lbrace \matr{Y}_{(3)}'\boxdot\left[\matr{X}_{(3)}'\right]^{-\beta'}\right\rbrace \left(\matr{B}'\khatri\matr{A}' \right)\\&\qquad\qquad\qquad\qquad+\frac{1}{\phi_{c}}\left\lbrace\matr{C}\boxdot\left[\matr{C}' \right]^{1-\beta_{c}}  \right\rbrace \\
\nabla\Upsilon_{\matr{C}'}^{+}=&\frac{1}{\phi'}\left[\matr{X}_{(3)}'\right]^{1-\beta'}\left(\matr{B}'\khatri\matr{A}' \right)+  \frac{1}{\phi_{c}}\left[\matr{C}'\right]^{1-\beta_{c}}.
\end{aligned}
\label{eq_grad_C_p}
\end{equation}\normalsize

\section{SNR for some coupling scenarios}
\label{app_2}
In this appendix, we present the expressions for the SNR of the tensors $ \tens{Y} $ and $ \tens{Y}' $ considered in simulations (Sec. \ref{simulations-sec}) following the hybrid Gaussian models. We present the SNR for the direct coupling model with unnormalized and normalized factors (Subsec. \ref{similarFactors-sec} and \ref{compressedCoupledCP-sec}) and for different samplings of a sum of sine functions (Subsec. \ref{differentSampling-sec}). 
\paragraph{Direct coupling}we recall briefly the simulation conditions: factors $ \matr{A} $, $ \matr{B} $, $ \matr{A}' $, $ \matr{B}' $ and $ \matr{C}' $ are drawn independently according to $ \mathcal{N}(0,1) $, the noise tensors $ \tens{V} $ and $ \tens{V}' $ are drawn from i.i.d. $\mathcal{N}(0,\sigma_n^2)$ and $\mathcal{N}(0,\sigma_n'^2)$ respectively. The first rows of $\matr{A}$, $\matr{B}$, $\matr{A}'$ and $\matr{B}'$ are set to $ 1 $ in the unnormalized case, while their columns are divided by their respective $\ell_2$ norms in the normalized cases. The elements $C_{i,j}$ are drawn from $ \mathcal{N}(C'_{i,j},\sigma_{c}^{2}) $.

In the unnormalized case, the SNR for tensor $\tens{Y}$ is written as follows:
\begin{eqnarray}
\text{SNR}(\tens{Y}) &= 10\log_{10}\left(\mathds{E}\left[\frac{\|\tens{X}\|_F^2}{\|\tens{Y}-\tens{X}\|_F^2}\right]\right) \\
                               &= 10\log_{10}\left(\frac{R(1+\sigma_c^2)}{\sigma_n^2} \right)
\label{SNR1}
\end{eqnarray}
while tensor $\tens{Y}'$ has a more usual SNR given by:
\begin{equation}
\text{SNR}(\tens{Y}') = 10\log_{10}\left(\frac{R}{\sigma_n^2}\right).
\label{SNR2}
\end{equation}

In the normalized case we have
\begin{equation}
\text{SNR}(\tens{Y}) = 10\log_{10}\left(\frac{R(1+\sigma_c^2)}{IJ\sigma_n^2} \right)
\label{SNR3}
\end{equation}
and
\begin{equation}
\text{SNR}(\tens{Y}') = 10\log_{10}\left(\frac{R}{I'J'\sigma_n^2}\right).
\label{SNR4}
\end{equation}
Note that for $ r<R $ shared components, as the model considered in Subsec. \ref{sharedComponent-sec}, $ \text{SNR}(\tens{Y}) $ changes to $ 10\log_{10}\left[(r\sigma_c^2+R)/(IJ\sigma_n^2) \right] $.
\paragraph{Sum of sines}in Subsec. \ref{differentSampling-sec} factors $ \matr{A} $, $ \matr{B} $, $ \matr{A}' $ and $ \matr{B}' $ are generated in the same way as in the previous paragraph, however the columns $ \vect{c}_{r} $ and $ \vect{c}'_{r} $ are generated by sampling with sampling periods $ T $ and $ T' $ the continuous functions $ c_{r}(t)=\sum_{i=1}^{N_{c}} \alpha_{ir}\sin(2\pi f_{i} t) $ with i.i.d. $ \alpha_{ir} $ drawn from $ \mathcal{N}(0,1) $. The SNR of the tensor models is
\begin{equation}\small
\text{SNR}(\tens{Y}) = 10\log_{10}\left[\frac{R\sum_{i=1}^{N_{c}}\sum_{k=1}^{K} \sin^2(2\pi f_{i} kT)}{IJK\sigma_n^2}\right].
\label{SNR5}
\end{equation}\normalsize




\section*{Acknowledgment}
We thank the anonymous reviewers and the associate editor for their careful reading and  helpful comments.


\ifCLASSOPTIONcaptionsoff
  \newpage
\fi

\bibliographystyle{IEEEtran}
\IEEEtriggeratref{31} 
\bibliography{farias_multimodal_fusion_references.bib}

\begin{thebibliography}{10}
\providecommand{\url}[1]{#1}
\csname url@samestyle\endcsname
\providecommand{\newblock}{\relax}
\providecommand{\bibinfo}[2]{#2}
\providecommand{\BIBentrySTDinterwordspacing}{\spaceskip=0pt\relax}
\providecommand{\BIBentryALTinterwordstretchfactor}{4}
\providecommand{\BIBentryALTinterwordspacing}{\spaceskip=\fontdimen2\font plus
\BIBentryALTinterwordstretchfactor\fontdimen3\font minus
  \fontdimen4\font\relax}
\providecommand{\BIBforeignlanguage}[2]{{%
\expandafter\ifx\csname l@#1\endcsname\relax
\typeout{** WARNING: IEEEtran.bst: No hyphenation pattern has been}%
\typeout{** loaded for the language `#1'. Using the pattern for}%
\typeout{** the default language instead.}%
\else
\language=\csname l@#1\endcsname
\fi
#2}}
\providecommand{\BIBdecl}{\relax}
\BIBdecl

\bibitem{Farias2015}
R.~Cabral~Farias, J.~E. Cohen, C.~Jutten, and P.~Comon, ``Joint decompositions
  with flexible couplings,'' in \emph{Latent Variable Analysis and Signal
  Separation}.\hskip 1em plus 0.5em minus 0.4em\relax Springer, 2015, pp.
  119--126.

\bibitem{Sui2012}
J.~Sui, T.~Adal{\i}, Q.~Yu, J.~Chen, and V.~Calhoun, ``A review of multivariate
  methods for multimodal fusion of brain imaging data,'' \emph{J. Neurosci.
  Meth.}, vol. 204, no.~1, pp. 68--81, 2012.

\bibitem{Acar2013a}
E.~Acar, A.~Lawaetz, M.~Rasmussen, and R.~Bro, ``Structure-revealing data
  fusion model with applications in metabolomics,'' in \emph{Conf. Proc. IEEE
  Eng. Med. Biol. Soc.}\hskip 1em plus 0.5em minus 0.4em\relax IEEE, 2013, pp.
  6023--6026.

\bibitem{Ermics2012}
B.~Ermi{\c{s}}, E.~Acar, and A.~T. Cemgil, ``Link prediction via generalized
  coupled tensor factorisation,'' \emph{arXiv preprint arXiv:1208.6231}, 2012.

\bibitem{Hotelling1936}
H.~Hotelling, ``Relations between two sets of variates,'' \emph{Biometrika},
  vol.~28, no. 3/4, pp. 321--377, 1936.

\bibitem{Bach2005}
F.~Bach and M.~Jordan, ``A probabilistic interpretation of canonical
  correlation analysis,'' Univ. of California, Berkeley, Tech. Rep. 688, 2005.

\bibitem{Harshman1984}
R.~A. Harshman and M.~E. Lundy, ``{D}ata preprocessing and the extended
  {P}{A}{R}{A}{F}{A}{C} model,'' in \emph{{R}esearch methods for multimode data
  analysis.}, H.~G. Law, C.~W. {Snyder Jr}, J.~A. Hattie, and R.~P. McDonald,
  Eds.\hskip 1em plus 0.5em minus 0.4em\relax Praeger, 1984, pp. 216--284.

\bibitem{Singh2008}
A.~Singh and G.~Gordon, ``Relational learning via collective matrix
  factorization,'' in \emph{Proc. 14th ACM SIGKDD Conf.}\hskip 1em plus 0.5em
  minus 0.4em\relax ACM, 2008, pp. 650--658.

\bibitem{Banerjee2007}
A.~Banerjee, S.~Basu, and S.~Merugu, ``Multi-way clustering on relation
  graphs,'' in \emph{Proc. SIAM Int. Conf. Data Mining}, 2007, pp. 145--156.

\bibitem{Lin2009}
Y.-R. Lin, J.~Sun, P.~Castro, R.~Konuru, H.~Sundaram, and A.~Kelliher,
  ``Meta{F}ac: community discovery via relational hypergraph factorization,''
  in \emph{Proc. ACM SIGKDD Conf.}\hskip 1em plus 0.5em minus 0.4em\relax ACM,
  2009, pp. 527--536.

\bibitem{Acar2013b}
E.~Acar, M.~Rasmussen, F.~Savorani, T.~N{\ae}s, and R.~Bro, ``Understanding
  data fusion within the framework of coupled matrix and tensor
  factorizations,'' \emph{Chemometr. Intell. Lab.}, vol. 129, pp. 53--63, 2013.

\bibitem{Acar2011}
E.~Acar, T.~G. Kolda, and D.~M. Dunlavy, ``All-at-once optimization for coupled
  matrix and tensor factorizations,'' \emph{arXiv preprint arXiv:1105.3422},
  2011.

\bibitem{Sorber2015}
L.~Sorber, M.~Van~Barel, and L.~De~Lathauwer, ``Structured data fusion,''
  \emph{IEEE J. Sel. Topics Signal Process.}, vol.~9, no.~4, pp. 586--600, June
  2015.

\bibitem{Sorensen2015a}
M.~S{\o}rensen and L.~De~Lathauwer, ``Coupled canonical polyadic decompositions
  and (coupled) decompositions in multilinear rank-$({L}_{r,n},{L}_{r,n}, 1)$
  terms. {P}art {I}: {U}niqueness,'' \emph{SIAM J. Matrix Anal. Appl.},
  vol.~36, no.~2, pp. 496--522, 2015.

\bibitem{Sorensen2015b}
M.~S{\o}rensen, I.~Domanov, and L.~De~Lathauwer, ``Coupled canonical polyadic
  decompositions and (coupled) decompositions in multilinear rank-
  $(l_{r,n},l_{r,n},1)$ terms -- part {II}: Algorithms,'' \emph{SIAM J. Matrix
  Anal. Appl.}, vol.~36, no.~3, pp. 1015--1045, 2015.

\bibitem{Yilmaz2011}
K.~Y{\i}lmaz, A.~Cemgil, and U.~Simsekli, ``Generalised coupled tensor
  factorisation,'' in \emph{Adv. Neural. Inf. Process. Syst.}, 2011, pp.
  2151--2159.

\bibitem{Csimcsekli2015}
U.~{\c{S}}im{\c{s}}ekli, T.~Virtanen, and A.~Cemgil, ``Non-negative tensor
  factorization models for {B}ayesian audio processing,'' \emph{Digit. Signal
  Process.}, 2015.

\bibitem{Seichepine2014}
N.~Seichepine, S.~Essid, C.~Fevotte, and O.~Capp\'e, ``Soft nonnegative matrix
  co-factorization,'' \emph{IEEE Trans. Sig. Process.}, vol.~62, no.~22, pp.
  5940--5949, Nov 2014.

\bibitem{Rivet2015}
B.~Rivet, M.~Duda, A.~Gu{\'e}rin-Dugu{\'e}, C.~Jutten, and P.~Comon,
  ``Multimodal approach to estimate the ocular movements during {EEG}
  recordings: a coupled tensor factorization method,'' in \emph{Int. Conf. of
  the IEEE Engineering in Medicine and Biology Society (EMBC)}, 2015.

\bibitem{Liu2001}
X.~Liu and N.~Sidiropoulos, ``Cram\'er-{R}ao lower bounds for low-rank
  decomposition of multidimensional arrays,'' \emph{IEEE Trans. Signal
  Process.}, vol.~49, no.~9, pp. 2074--2086, 2001.

\bibitem{TichPK13:tsp}
P.~Tichavsky, A.~H. Phan, and Z.~Koldovsky, ``Cramer-{R}ao-induced bounds for
  {CANDECOMP}/{PARAFAC} tensor decomposition,'' \emph{IEEE Trans. Sig. Proc.},
  vol.~61, no.~8, pp. 1986--1997, Apr. 2013.

\bibitem{Yokoya2012}
N.~Yokoya, T.~Yairi, and A.~Iwasaki, ``Coupled nonnegative matrix factorization
  unmixing for hyperspectral and multispectral data fusion,'' \emph{IEEE Trans.
  Geosci. Remote Sens.}, vol.~50, no.~2, pp. 528--537, 2012.

\bibitem{Brew78:cs}
J.~W. Brewer, ``Kronecker products and matrix calculus in system theory,''
  \emph{IEEE Trans. on Circuits and Systems}, vol.~25, no.~9, pp. 114--122,
  Sep. 1978.

\bibitem{Kolda2009}
T.~Kolda and B.~Bader, ``Tensor decompositions and applications,'' \emph{SIAM
  {R}ev.}, vol.~51, no.~3, pp. 455--500, 2009.

\bibitem{Comon2014}
P.~Comon, ``Tensors: {A} brief introduction,'' \emph{IEEE Sig. Process. Mag.},
  vol.~31, no.~3, pp. 44--53, 2014.

\bibitem{Comon2009}
P.~Comon, X.~Luciani, and A.~de~Almeida, ``Tensor decompositions, alternating
  least squares and other tales,'' \emph{J. Chemometrics}, vol.~23, no. 7-8,
  pp. 393--405, 2009.

\bibitem{Rogers2013}
\BIBentryALTinterwordspacing
M.~Rogers, L.~Li, and S.~J. Russell, ``Multilinear dynamical systems for tensor
  time series,'' in \emph{Advances in Neural Information Processing Systems
  26}, C.~Burges, L.~Bottou, M.~Welling, Z.~Ghahramani, and K.~Weinberger,
  Eds., 2013, pp. 2634--2642. [Online]. Available:
  \url{http://papers.nips.cc/paper/5117-multilinear-dynamical-systems-for-tens%
or-time-series.pdf}
\BIBentrySTDinterwordspacing

\bibitem{Aldroubi2001}
A.~Aldroubi and K.~Gr{\"o}chenig, ``Nonuniform sampling and reconstruction in
  shift-invariant spaces,'' \emph{SIAM Rev.}, vol.~43, no.~4, pp. 585--620,
  2001.

\bibitem{Jorgensen1992}
B.~J{\o}rgensen, \emph{The Theory of Exponential Dispersion Models and Analysis
  of Deviance}, ser. Monographs in mathematics.\hskip 1em plus 0.5em minus
  0.4em\relax {CNPQ}, {IMPA} (Brazil), 1992, no.~51.

\bibitem{Yilmaz2012}
Y.~Yilmaz and A.~Cemgil, ``Alpha/beta divergences and tweedie models,''
  \emph{arXiv preprint arXiv:1209.4280}, 2012.

\bibitem{DattS91:laa}
B.~N. Datta and Y.~Saad, ``Arnoldi methods for large {S}ylvester-like observer
  matrix equations, and an associated algorithm for partial spectrum
  assignment,'' \emph{Lin. Alg. Appl.}, vol. 154-156, no.~2, pp. 225--244,
  1991.

\bibitem{GoluNV79:tac}
H.~Golub, S.~Nash, and C.~V. Loan, ``A {H}essenberg-{S}chur method for the
  problem {AX + XB = C},'' \emph{IEEE Trans. Auto. Control}, vol.~24, 1979.

\bibitem{Bro1998}
R.~Bro, ``\textit{Multi-way Analysis in the Food Industry: Models, Algorithms,
  and Applications},'' Ph.D. dissertation, University of Amsterdam, The
  Netherlands, 1998.

\bibitem{Munkres1957}
J.~Munkres, ``Algorithms for the assignment and transportation problems,''
  \emph{Journal of the Society for Industrial and Applied Mathematics}, vol.~5,
  no.~1, pp. 32--38, 1957.

\bibitem{DeLathawer2000}
L.~De~Lathauwer, B.~De~Moor, and J.~Vandewalle, ``A multilinear singular value
  decomposition,'' \emph{SIAM J. Matrix Anal. Appl.}, vol.~21, no.~4, pp.
  1253--1278, 2000.

\bibitem{CullW85}
J.~Cullum and R.~Willoughby, \emph{Lanczos Algorithms}.\hskip 1em plus 0.5em
  minus 0.4em\relax Birkhauser, 1985, vol.~I.

\bibitem{Halko2011}
N.~Halko, P.-G. Martinsson, and J.~Tropp, ``Finding structure with randomness:
  Probabilistic algorithms for constructing approximate matrix
  decompositions,'' \emph{SIAM rev.}, vol.~53, no.~2, pp. 217--288, 2011.

\bibitem{Lee2001}
D.~Lee and H.~Seung, ``Algorithms for non-negative matrix factorization,'' in
  \emph{{A}dvances in {N}eural {I}nformation {P}rocessing {S}ystems}, 2001, pp.
  556--562.

\bibitem{Fevotte2011}
C.~F{\'e}votte and J.~Idier, ``Algorithms for nonnegative matrix factorization
  with the $\beta$-divergence,'' \emph{Neural Comput.}, vol.~23, no.~9, pp.
  2421--2456, 2011.

\bibitem{VanTrees2007}
H.~Van~Trees and K.~Bell, \emph{Bayesian Bounds for Parameter Estimation and
  Nonlinear Filtering/Tracking}.\hskip 1em plus 0.5em minus 0.4em\relax
  Wiley-IEEE Press, 2007.

\bibitem{Margolis2008}
E.~Margolis and Y.~Eldar, ``Nonuniform sampling of periodic bandlimited
  signals,'' \emph{IEEE Trans. Sig. Process.}, vol.~56, no.~7, pp. 2728--2745,
  2008.

\end{thebibliography}

~\vfill


\end{document}